\documentclass[5p,twocolumn,11pt]{elsarticle}
\usepackage{amsbsy}
\usepackage{amssymb}
\usepackage{amsmath}
\usepackage{graphicx}
\usepackage{color}
\usepackage{xcolor}
\usepackage{natbib}
\usepackage{hyperref}

\newcommand{\be}{\begin{equation}}
\newcommand{\ee}{\end{equation}}
\newcommand{\bear}{\begin{eqnarray}}
\newcommand{\eear}{\end{eqnarray}}

\newcommand{\derparn}[2] {\frac{\partial #2}{\partial #1}}

\journal{Computer Physics Communications}
\begin{document}

\begin{frontmatter}

\title{A Simflowny-based high-performance 3D code for the generalized induction equation}

\author[label1,label2]{Daniele Vigan\`o}
\author[label1,label2]{David Mart\'inez-G\'omez}
\author[label3]{Jos\'e A. Pons}
\author[label1,label2]{Carlos Palenzuela}
\author[label1,label2]{Federico Carrasco}
\author[label1,label2]{Borja Mi\~{n}ano}
\author[label1,label2]{Antoni Arbona}
\author[label1,label2]{Carles Bona}
\author[label1,label2]{Joan Mass\'o}


\address[label1]{Departament  de  F\'{\i}sica, Universitat  de  les  Illes  Balears  and  Institut  d'Estudis Espacials  de  Catalunya,  Palma  de  Mallorca, Baleares  E-07122,  Spain}
\address[label2]{Institut d'Aplicacions Computacionals de Codi Comunitari (IAC3),  Universitat  de  les  Illes  Balears,  Palma  de  Mallorca,  Baleares  E-07122,  Spain}
\address[label3]{Departament de F\'{\i}sica Aplicada, Universitat d'Alacant, Ap. Correus 99, 03080 Alacant, Spain}


\begin{abstract}
In the interior of neutron stars, the induction equation regulates the long-term evolution of the magnetic fields by means of resistivity, Hall dynamics and ambipolar diffusion. Despite the apparent simplicity and compactness of the equation, the dynamics it describes is not trivial and its understanding relies on accurate numerical simulations. While a few works in 2D have reached a mature stage and a consensus on the general dynamics at least for some simple initial data, only few attempts have been performed in 3D, due to the computational costs and the need for a proper numerical treatment of the intrinsic non-linearity of the equation. Here, we carefully analyze the general induction equation, studying its characteristic structure, and we present a new Cartesian 3D code, generated by the user-friendly, publicly available {\em Simflowny} platform. The code uses high-order numerical schemes for the time and spatial discretization, and relies on the highly-scalable {\em SAMRAI} architecture for the adaptive mesh refinement. We present the application of the code to several benchmark tests, showing the high order of convergence and accuracy achieved and the capabilities in terms of magnetic shock resolution and three-dimensionality. This paper paves the way for the applications to a realistic, 3D long-term evolution of neutron stars interior and, possibly, of other astrophysical sources.
\end{abstract} 

\begin{keyword}
MHD; neutron stars; Simflowny; magnetic fields; high resolution shock capturing; AMR
\end{keyword}
\end{frontmatter}


\section{Introduction}
The induction equation plays a key role in many physical problems involving magnetized plasma flows. In the most common and simple model for the study of plasmas, namely ideal magnetohydrodynamics (MHD), this equation arises from the combination of Faraday's law and the ideal Ohm's law for a perfect conductor, which simply relates the electric field to the cross product of the fluid velocity and the magnetic field. 
The two basic assumptions in {\it ideal MHD} are that all the components of the plasma are strongly coupled (a single fluid),  and that the resistive timescales are much longer than the dynamical timescales of interest. If some of the previous assumptions are relaxed, more complex theoretical descriptions are required. For example, relaxing the assumption of negligible resistivity leads to {\it resistive MHD}, in which a term proportional to the electric current, is introduced in Ohm's law.

In addition, each constituent of the plasma may have a different hydrodynamical velocity. Thus, multi-fluid models with a separate set of equations for each component would be required to properly describe the dynamics of the plasma \citep{schunk77}. Although they are more accurate and have a larger range of applicability, their numerical cost is
significant. Nevertheless, the additional complexity can often be synthetized in simpler descriptions that combine the equations of all, or several of, the constituents of the plasma and rely on the use of a generalized Ohm's law.

An important  MHD extension of practical interest is that of {\it Hall MHD}, a  limit in which one charged component (usually electrons) is free to move, while the other charged particles (typically ions) have a restricted mobility, or in other words, their gyro-frequency is much longer than other relevant timescales. Under this condition, the electromagnetic field evolves according to a non-linear induction equation with an electric field corrected by an extra term proportional to the Lorentz force. 
Hall MHD finds direct applications in plasma physics and astrophysics \citep{witalis86}. For instance, it is used to describe opening switches in plasma laboratories, magnetic reconnections in the magneto-pause and magneto-tail (see, e.g. \cite{deng2001,mozer2002,bard18}), interstellar magnetic field dynamics, plasma expansion models, small-scale dynamos, and proto-planetary disks \cite{kunz2004,pandey2008,bethune16}.
The limit in which ions are considered completely fixed, and only the dynamics of one charged fluid (electrons) is relevant is also known as {\it electron MHD} (eMHD). A comprehensive review of Hall MHD, including its characteristic modes and the most suitable numerical methods to deal with it is described in \cite{huba91,huba03}. 

A different approximation holds if we assume that both positive and negatively charged fluids are strongly coupled to each other, but they are weakly coupled to the neutral components, so that the difference of their velocities is smaller than their average (but not zero). In this case, there is a drift of the coupled charged components with respect to the neutral fluid. This is usually called ambipolar diffusion, and has been studied in a variety of astrophysical scenarios, like solar physics \citep{khomenko2012,soler15}, star formation \citep{fiedler92},  accretion disks and molecular clouds \citep{jones11,gressel2015}.

Focusing on neutron stars, all the above described effects have been found to be important at different stages of their long-term evolution (see eg. \cite{goldreich92}).
These compact stellar objects are thought to be constituted by a highly (or very likely super-) conductive core with a radius of about 10 km, where matter is extremely packed under unique and only partially understood conditions of high pressure and density for which the nuclear forces dominate; a $\sim$ km-thick crust, resembling an ultra-high-density ($10^{10}-10^{14}$ g cm$^{-3}$) metal; a thin liquid and/or gaseous envelope (sometimes called ocean), and usually a $\sim$cm-thick atmosphere. In the crust, fully pressure-ionized electrons are free to move in a lattice of heavy ions, therefore being well represented by the eMHD limit. The evolution of magnetic fields in the crust of neutron stars has been thoroughly studied \cite{hollerbach02, hollerbach04,pons07b,pons09,vigano12a,vigano13,geppert14} with more or less realistic 2D simulations over relevant timescales (typically, Myr), including the resistive (or Ohmic) and Hall terms and sometimes coupling temperature and magnetic field evolution. Independently, a set of papers \cite{gourgouliatos13,gourgouliatos14a,gourgouliatos14b} have confirmed the general picture of the Hall dynamics found in the works mentioned above. Those works used a code that includes only the magnetic field evolution and pointed out new interesting effects. They find a Hall-attractor solution, which corresponds to a quasi-stationary state that slowly evolves due to the simultaneous Hall and resistive terms. Recently, the first 3D attempts to the whole problem have been presented in \cite{wood2015} and \cite{gourgouliatos18}, in which the authors use a mixed spectral-finite difference code to simulate a neutron star crust coupled with external vacuum. Their results show new dynamics and the creation and persistence of km-size magnetic structures over long timescales.
In the core, however, the situation is less clear and there is increasing evidence that ambipolar diffusion and/or fluid motions are key to describe the magnetic field evolution. Several recent papers have started to address this problem with numerical simulations or novel theoretical ideas \cite{elfritz16,castillo17,gusakov17,passamonti17a,passamonti17b,kantor18,bransgrove18,ofengeim18}.

The long-term evolution of magnetic fields, given by the apparently simple induction equation, is an excellent benchmark to test several theoretical issues. These are related to the interdependence with the evolving temperature, the composition and the state of the core (which affects the relative importance of one or another term and the dynamical timescales), the possible presence of nuclear pasta phase in the crust/core transition layer, the chemical composition of the atmosphere, the initial data (the magnetic topology at birth is basically unknown), and the treatment of the boundary condition at the star's surface. Furthermore, numerical issues, regarding the non-linearity of the Hall and ambipolar terms, have represented a major technical obstacle to a systematic study of the various scenarios. The three-dimensionality is thought to be a crucial factor to study realistic cases (in analogy to solar magnetism) and, ultimately, to compare the theoretical predictions on the observables with the observations, especially for magnetars, which are young and strongly magnetized neutron stars. Thus, the need for High-Performance-Computing (HPC) is clear. For this reason, in this paper we present a numerical code and a series of testbed simulations of a generalized form of the induction equation. This code was built by using {\em Simflowny}~\citep{arbona13,arbona18}, a versatile platform able to automatically generate parallelized codes for partial differential equations. It employs the Adaptive Mesh Refinement (AMR) libraries of {\em SAMRAI} \citep{hornung02}, and a graphical user interface which easily allows to implement equations and to choose among different time and space discretization schemes. The choice of {\em SAMRAI} is mainly due to its proven high scalability \citep{gunney16}, which is an important requirement to deal with 3D simulations of non-linear dynamics. The paper is structured as follows. In \S \ref{sec:generalized_induction} we briefly review the mathematical form of the induction equation under different assumptions, and its characteristic structure. In \S \ref{sec:simflowny} we summarize the main features of {\em Simflowny}.  In  \S \ref{sec:tests} and \ref{sec:3d} we present a number of tests and toy models of the different regimes in 2D and 3D. Finally, we discuss the capability of the platform to handle the problem, its current limitations, and we give an overview of future applications and extensions.

\section{Generalized induction equation}\label{sec:generalized_induction}

\subsection{Induction equation}\label{sec:induction}
Faraday's law, which describes the temporal evolution of the magnetic field in terms of the spatial variations of the electric field, is given by (in SI units):
\begin{equation}
\frac{\partial \vec{B}}{\partial t} = - \vec{\nabla}\times \vec{E}.
\end{equation}

In a fluid description of plasmas, Faraday's law is accompanied by a set of hydrodynamic equations, Amp\`ere's law, and Ohm's law. The latter allows to close the system of equations by relating the electric field to other variables such as velocities and the magnetic field. Solving this full system of equations can be complex and, in most situations, one must assume some simplifications.
In the simplest case, when matter is described as a single fluid, the electric field in the reference frame comoving with matter is simply related to the electrical conductivity, $\sigma$, and the electrical current density, $\vec{j}$, by 
\begin{equation}
\vec{j} = \sigma \vec{E},
\end{equation}
which, together with Ampere's law (neglecting the displacement currents)
\begin{equation}\label{eq:current_mhd}
\vec{j} = \vec{\nabla}\times \vec{B},
\end{equation}
and Faraday's law, results in an induction equation that adopts the form of a vectorial diffusion equation:
\begin{equation}
\frac{\partial \vec{B}}{\partial t} +  \vec{\nabla}\times \left( \eta \vec{\nabla}\times \vec{B} \right) = 0,
\end{equation}
where we have defined the magnetic diffusivity $\eta= \frac{1}{\sigma}$.

In a more general case, the situation is more complex and different terms arise, as described in detail in \citep{goldreich92} for neutron stars. However, we can build a general induction equation to study its mathematical properties and possible strategies for numerical solutions. The electric field, as any other vector, can be expressed in any convenient basis. Let us write a general form as
\begin{equation}\label{eq:efield_general}
\vec{E} = f_d \vec{j} + f_h (\vec{j}  \times \vec{B}) - f_a  (\vec{j}  \times \vec{B})\times \vec{B}
\end{equation}

Therefore, the general vectorial equation that we aim to solve is
\begin{equation}\label{eq:ind_general}
\frac{\partial \vec{B}}{\partial t} +  \vec{\nabla}\times \left( f_d \vec{j} + f_h (\vec{j}\times \vec{B}) - f_a (\vec{j}  \times \vec{B})\times \vec{B} \right) = 0
\end{equation}
with the current given by eq.~(\ref{eq:current_mhd}).

Physically, the coefficients $f_d=4\pi\eta/c^2$ and $f_a$ are determined from collision terms in the Euler equations and are positive-defined, while $f_h$ depends on the charge density of the charges carrying the current: it is positive if negative charges are faster than the positive charges (as in the eMHD limit), and negative otherwise. In general, all these coefficients depend on the physical conditions (density, temperature, chemical composition), and one or another term may dominate in different regimes.
Other possible terms contributing to the electric field are due to the thermo-electric effect \citep{geppert91} and the chemical potential imbalance \citep{goldreich92}, wherever the gradient of temperature and of chemical potential, respectively, induce a movement of the charges. Here we do not consider such terms, since they require the coupled evolution with the temperature and the chemical composition.

Let us analyze each term appearing in the electric field equation. The first term on the r.h.s. in eq.~(\ref{eq:efield_general}) is the purely resistive (Ohmic) term. The  second and third terms (the {\it Hall term} and the {\it ambipolar diffusion} term) can be combined in a vectorial advection term ($\vec{v} \times \vec{B}$), in which $\vec{v}= -f_h \vec{j} + f_a (\vec{j}  \times \vec{B})$ is interpreted as a weighted average velocity of the charged fluid that implicitly depends on $B$.

The Hall term is proportional to the current, it is quadratic in $\vec{B}$, and hides a Burgers-like behavior, leading to the creation of localized current sheets (see e.g. \cite{vainshtein00} and test \S~\ref{sec:hall_nl}). This term conserves the magnetic energy by definition, being able only to redistribute it in different scales without dissipation per se. One can easily probe that it conserves the magnetic energy (by taking the scalar product of the equation by $\vec{B}$), except for surface terms (Poynting flux), or noting that the dissipated energy is $\vec{E}\cdot\vec{j}\propto\vec{j}\cdot(\vec{j}\times\vec{B})=0$.

The ambipolar diffusion term, which is cubic in $\vec{B}$, is related to a velocity with the same direction of the total magnetic (Lorentz) force acting on the charged particles, given by $(\vec{j} \times \vec{B})$. Its effect is to dissipate the currents perpendicular to $\vec{B}$, having no effect in the current flowing along magnetic field lines. In other words, this term acts to align magnetic and current fields, thus bringing the system into a force-free configuration, characterized by definition by $\vec{j} \times \vec{B}=0$. Note that a formally similar term has been used to find configurations of twisted force-free magnetosphere, under the name of magneto-frictional term \citep{roumeliotis94,vigano11}. In this paper, we will present a simple application in \S~\ref{sec:poloidal} to treat the external part of the star.

Ideally, one wants to be able to numerically solve the induction equation for arbitrary values of all these parameters. But before entering into numerical details, it is worth analyzing the characteristic structure of the system.

\subsection{Characteristic structure}\label{sec:characteristic}

The characteristic analysis captures the behavior of perturbations around a background field at a particular point in space (and time), generally related with the study of local stability for the given evolution system and the concept of well-posedness. 
Moreover, it might also provide useful information for numerical applications, in particular how to impose boundary conditions or to treat interfaces. With this purpose, let us consider perturbations of equation \eqref{eq:ind_general} over a smooth background solution $\vec{B}_o$. We shall look then for plane-wave solutions of the form, 
\begin{equation}
\vec{B} = \vec{B}_o + \vec{B}_1 \, e^{i(\vec{k}\cdot \vec{x} - \omega t)}
\end{equation}
and take the high-frequency limit, in which neither the amplitude $\vec{B}_1$ nor the coefficients ($f_d, f_{h}, f_{a}$) would vary in space and it has been assumed that $|\vec{B}_1 | \ll |\vec{B}_o |$. Hence, 
\begin{equation}
\frac{\partial}{\partial t} \rightarrow -i\omega \quad \text{;} \quad  \vec{\nabla} \rightarrow i \vec{k} 
\end{equation}
Before continuing, let us introduce first some convenient notational abbreviations:
\begin{equation}
 A_k := \hat{k}\cdot \vec{A} \text{;} \quad \vec{A}_{p} := \vec{A} - A_k \, \hat{k} \text{;} \quad \vec{A}_{q} =  \hat{k} \times \vec{A}  \nonumber
\end{equation}
for any vector $\vec{A}$. Notice $\vec{A}_{p}$ and $\vec{A}_{q}$ are both orthogonal to the propagation direction $\hat{k}:= \vec{k}/k$ (with $k\equiv|\vec{k}|$) and also perpendicular among them. With this notation:
\begin{eqnarray*}
 \vec{\nabla}\times \vec{B} &\rightarrow& i \, k\, \vec{B}_{1q} \\
 (\vec{\nabla}\times \vec{B})\times \vec{B} &\rightarrow& i \, k\, (\vec{B}_{1q} \times \vec{B}_o ) \\
 \left( (\vec{\nabla}\times \vec{B})\times \vec{B}\right) \times \vec{B} &\rightarrow& i k\left[  (\vec{B}_o \cdot \vec{B}_{1q} ) \vec{B}_o - B_{o}^2 \vec{B}_{1q}  \right] ~.
\end{eqnarray*}
Therefore, the terms in eq.~(\ref{eq:ind_general}) expand as follows:
\begin{eqnarray*}
 \vec{\nabla}\times \vec{j} &\rightarrow&  k^2 \vec{B}_{1p} \\
 \vec{\nabla}\times \left(  \vec{j}\times \vec{B} \right)  &\rightarrow& - k^2 B_{ok} \vec{B}_{1q}\\
 \vec{\nabla}\times \left[ \left( \vec{j}\times \vec{B}\right) \times \vec{B} \right]  &\rightarrow& - k^2 \left[  (\vec{B}_o \cdot \vec{B}_{1q} ) \vec{B}_{oq} + B_{o}^2 \vec{B}_{1p}  \right] 
\end{eqnarray*}
and the linearized system finally reads:
\begin{eqnarray}\label{linearized-system}
 && i \frac{\omega}{k^2} \vec{B}_1 = \left( f_d + f_a B_{o}^2 \right) \vec{B}_{1p} + \nonumber\\
 &&  - f_h B_{ok} \vec{B}_{1q} + f_a (\vec{B}_o \cdot \vec{B}_{1q} ) \vec{B}_{oq} 
\end{eqnarray}
Contracting the equation above  with $\vec{k}$ reveals that there is a zero-mode ($\omega = 0 $) associated with the solenoidal constraint $\vec{\nabla}\cdot \vec{B} = 0$, 
with its eigenvector along the propagation direction, $\vec{B}_{1} \propto \vec{k} $. 
This might be interpreted by saying that any deviation from the physical constraint $\vec{\nabla}\cdot \vec{B}=0$ will not propagate. Thus, hereafter we focus only on the physical perturbations, which are transversal to $\vec{k}$. The eigenvalues are given by:
\begin{equation}\label{eigenvalues}
i \frac{\omega^{\pm}}{k^2} = f_d + \frac{1}{2} f_a (B_{o}^2 + B_{ok}^2 ) \pm \frac{1}{2}\sqrt{f_{a}^2 B_{op}^4 - 4 f_{h}^2 B_{ok}^2}
\end{equation}
with their respective eigenvectors~\footnote{Note that the particular case ($f_h = 0$) cannot be directly recovered from this expression. 
In this case, the eigenvectors must be obtained directly from eq.~(\ref{linearized-system}).},
\begin{equation}\label{eigenvectors}
\vec{B}_{1}^{\pm} =  2 B_{ok} \vec{B}_{op} - \left[ f_r B_{op}^2 \mp \sqrt{f_{r}^2 B_{op}^4 - 4 B_{ok}^2} \right] \vec{B}_{oq}
\end{equation}
where we have defined the ratio, $f_r := f_a / f_h$.

In the particular situation where $\vec{B}_o \parallel \vec{k}$ (i.e~$\vec{B}_{op} = 0$), the solution can be re-written as:
\begin{eqnarray}
 \vec{B}_{1}^{\pm} &=&  \hat{p}  \pm i  \hat{q} \\
 i \omega^{\pm} &=&   k^2 \left\lbrace f_d + f_a B_{ok}^2 \pm i f_{h} |B_{ok}| \right\rbrace 
\end{eqnarray}
being $\hat{p}$ and $\hat{q}$ any two transversal unit vectors satisfying $\hat{k} \times \hat{p} = sign(B_{ok}) \, \hat{q} $. 
These solutions represent the familiar whistler waves (see e.g.~\cite{huba03}), commonly seen for instance in the Earth's ionosphere \citep{helliwell65,nunn74}, triggered by high-frequency transients produced by natural events like lightnings. They have a practical importance for spacecraft safety and space weather.

Notice that in the general dispersion relation \eqref{eigenvalues}, the positive signs of the Ohmic and ambipolar diffusion coefficients result in the decay of the modes, guarantying local well-posedness. 
In the ideal case where both vanish (i.e.~$f_d = f_a = 0$), the system becomes hyperbolic and the characteristic modes are no longer damped. Namely,
\begin{eqnarray}
 \vec{B}_{1}^{\pm} &=&  \hat{p}  \pm i  \hat{q} \label{eq:whistler_ideal} \\
 i \omega^{\pm} &=&  \pm i k^2 f_{h} |B_{ok}| \label{eq:whistler_ideal2}
\end{eqnarray}

Another solution for the ideal case can be obtained by relaxing the assumption of high-frequency limit and considering a non-zero gradient of $f_h$. 
In this scenario, the so-called Hall drift wave is recovered (see e.g.~\cite{huba91}). 
This is a single mode that propagates along the direction $\nabla f_h \times \vec{B}_o$, with a frequency given by $\omega = k^2 |\nabla f_h \times \vec{B}_o |$.
See \ref{app:characteristic} for a general derivation.

Finally, in the case with $f_a=f_h=0$ and constant $f_d\neq 0$, the characteristic modes are purely dissipative and are given by any two transversal (linearly independent) vectors, with a frequency $i\omega=k^2 f_d$. Interestingly, the induction equation admits here global solutions such that $\vec{\nabla}\times\vec{B}=\mu\vec{B}$ (with $\mu$ constant), which decays in a self-similar way, i.e.
\begin{equation}\label{eq:selfsimilar_bessel}
\vec{B} = \vec{B}_{in}e^{-t/\tau_d} 
\end{equation}
with $\tau_d= 4\pi / f_d c^2 \mu^2$. When $f_d$ or $\mu$ varies with the position (which is the realistic case for a neutron star), the dissipation is not homogeneous, the field configuration is not self-similar, and the decay will be enhanced in the regions with highest values of $f_d$ and $\vec{\nabla}\times\vec{B}$.

\subsection{Divergence cleaning}

Previous works used numerical algorithms that preserve the $\vec{\nabla}\cdot\vec{B}=0$ constraint by construction, like the use of staggered grids for finite difference schemes \citep{toth00}. Since we employ a more general and flexible code with a standard grid (all components of the fields are defined and evolved at every grid node), this constraint is not guaranteed. Therefore, we adopt a divergence cleaning method~\citep{dedner02}, extensively used in MHD, consisting of the addition of a scalar field as follows:
\begin{eqnarray}
  && \frac{\partial \vec{B}}{\partial t} +  \vec{\nabla}\times \vec{E} + \vec{\nabla}\Phi = 0 \nonumber \\
  && \frac{\partial \Phi}{\partial t} + c_h^2 \vec{\nabla}\cdot\vec{B} = -\kappa \Phi
\end{eqnarray}
where $\Phi$ is a scalar field that allows the propagation and damping of divergence errors.
Notice that $c_h$ is the propagation speed of the constraint-violating modes $\Phi \neq 0$, which decay exponentially on a timescale $1/\kappa$.
We implement this equation in our code to ensure that the constraint is always preserved, except in the case where the treatment is not needed because the constraint $\vec{\nabla}\cdot\vec{B}=0$ is numerically guaranteed by construction (tests where only one component of $\vec{B}$ is non-zero in \S~\ref{sec:halldrift}, \S~\ref{sec:hall_nl}, \S~\ref{sec:ambipolar_test}). The optimal cleaning is reached for $c_h=\kappa$. However, if $\kappa$ is too large, the system of equations becomes stiff and it is difficult to evolve with explicit numerical schemes. We set $c_h=\kappa=4$, unless otherwise specified (see also \S~\ref{sec:poloidal} for a comparison of results of different values of these parameters).

In order to quantify the deviation from the constraint, one has to compare the normalized L2-norm of the divergence, ${\cal C}$, with the magnetic energy (setting the magnetic permeability $\mu_0=1$ for simplicity):

\begin{eqnarray}
 && {\cal C} = \int_V ||\vec{\nabla}\cdot \vec{B}||^2 (\Delta x)^2 {\rm d}V \label{eq:divb_integrated} \\
 && {\cal E}_m = \int_V \frac{B^2}{2} {\rm d}V
\end{eqnarray}
where, in the case of applying AMR, $\Delta x$ is the grid size corresponding to the finest resolution.

\section{Numerical methods}\label{sec:simflowny}

\subsection{Platform and infrastructure}

The code presented here has been generated by using {\em Simflowny}~\cite{arbona13,arbona18} together with the infrastructure {\em SAMRAI} \citep{hornung02,gunney16}. {\it Simflowny} is an open-source and user-friendly platform developed by the IAC3 group since 2008 to facilitate the use of HPC infrastructures to non-specialist scientists. It allows to easily implement scientific dynamical models, by means of a Domain Specific Language, and a web-based integrated development environment, which automatically generates efficient parallel code for simulation frameworks. {\it Simflowny} splits the physical models and problems from the numerical techniques. The automatic generation of the simulating code allows to properly include the parallelization features, which in this case rely on the {\em SAMRAI} infrastructure~\cite{hornung02}\footnote{See also the website\\{\tt https://computation.llnl.gov/project/SAMRAI/}}. {\em SAMRAI} is a patch-based structured AMR developed over more than 15 years by the Center for Applied Scientific Computing at the Lawrence Livermore National Laboratory. The latest upgrades on the AMR algorithms allow to improve the performance and reach a good scaling on up to 1.5M cores and 2M MPI tasks~\cite{gunney16}, at least for some specific problems. The combination of these two platforms provides a final code with a good balance of speed, accuracy, scalability, ability to switch physical models (flexibility), and the capacity to run in different infrastructures (portability).

\subsection{Discretization Schemes}

Our system of equations can be written formally in conservation law form, namely 
\begin{eqnarray}\label{PDEequationdecomposed}
\partial_t {\bf U} + \partial_k {\cal F}^k({\bf U}) = 0
\end{eqnarray}
where ${\bf U}=\{ \vec{B}, \Phi \}$ is the list of evolved fields and ${\cal F}^k({\bf U})$ their corresponding fluxes which might be non-linear but depend only on the fields and not on their derivatives. Notice then that this is not completely true in the eMHD system due to the presence in the fluxes of the current $\vec{j}$, defined by  Eq.(\ref{eq:current_mhd}).

The discretization of the continuum equations is performed by using the Method of Lines, which allows to address separately the time and the space discretization.
Space finite difference schemes based on Taylor expansions, which are extremely suitable for smooth solutions, might be a good option to calculate the current, which might be computed in two different ways: either by a direct finite difference discretization, or invoking the Stokes theorem to integrate the circulation of the magnetic field on a closed path. Interestingly, at second order accuracy both approaches lead to the same discretization scheme, which is our preferred choice here (see \ref{app:j} for a discussion about other choices), and are simply composed by terms like
\begin{equation}
D_x B^y_{i} = \frac{B^y_{i+1} - B^y_{i-1}}{2 \Delta x} \nonumber 
\end{equation}
where $i$ identifies a specific discretized cell position on the grid along a direction perpendicular.

However, these simple finite difference operators are not the optimal choice for the spatial discretization of fluxes in intrinsically non-linear systems like the eMHD. In that case, it is advisable to use High-Resolution-Shock-Capturing (HRSC) methods~\citep{toro97}, to deal with the possible appearance of shocks and to take advantage of the existence of weak solutions in the equations. We will therefore use a conservative scheme to discretize the fluxes, which in one dimension reads:
\begin{eqnarray}
\label{conservative_discretization}
\partial_t {\bf U}_{i} = - \frac{1}{\Delta x} \left({\bf F}^{x}_{i+1/2} - {\bf F}^{x}_{i-1/2}\right) 
\nonumber 
\end{eqnarray}
where ${\bf F}^{x}_{i\pm 1/2}$ are the set of fluxes along the $x$-direction evaluated at the interfaces between two neighboring cells, located at $x_{i\pm 1/2}$. The crucial issue in HRSC methods is how to approximately solve the Riemann problem, by reconstructing the fluxes at the interfaces such that no spurious oscillations appear in the solutions. This calculation consists in two steps:

\begin{itemize}
	\item We consider the following combination of the fluxes and the fields, at each node $i$:
	\begin{eqnarray}\label{flux_decomposition}
	   F^{\pm}_{i} = \frac{1}{2} \left( F_i \pm \lambda U_i \right) 
	\end{eqnarray}
	where $\lambda$ is the maximum propagation speed of the system in the neighboring points. Then, we reconstruct the flux at the left (right) of each interface, e.g. $F^L_{i+1/2}$ ($F^R_{i+1/2}$), using the values $\{F^+\}$ ($\{F^-\}$) from the neighboring nodes $\{x_{i-n},..,x_{i+1+n}\}$. The number $2(n+1)$ of such neighbors used in the reconstruction procedure depends on the order of the method. {\it Simflowny} already incorporates some commonly used reconstructions, like PPM~\cite{colella84} and MP5~\cite{suresh97}, as well as other implementations like the FDOC families~\cite{bona09} which are almost as fast as centered finite difference schemes at the cost of some bounded spikes near the shock regions. Here we mostly focused on the Weighted-Essentially-Non-Oscillatory (WENO) reconstructions~\cite{jiang96,shu98}, which is our preferred choice for their flexibility (i.e., they can achieve any order of accuracy) and robustness. The detailed implementation of the WENO flavors used here can be found in~\citep{palenzuela18}.

    \item We use a flux formula to compute the final flux at each interface, e.g.:
	\begin{equation}
	\label{LLF2}
   	 F_{i+1/2} = F^L_{i+1/2} + F^R_{i+1/2}
	\end{equation}
	Note that our reconstruction method does not require the characteristic decomposition of the system of equations (i.e., the full spectrum of characteristic velocities). At the lowest order reconstruction $F^L_{i+1/2}=F^+_i$ and $F^R_{i+1/2}=F^-_{i+1}$, so that the flux formula~(\ref{LLF2}) reduces to the popular and robust Local-Lax-Friedrichs flux~\cite{toro97}.

\end{itemize}

The time integration of the resulting semi-discrete equations is performed by using a $4^{\rm th}$-order Runge-Kutta scheme, which ensures the stability and convergence of the solution for a small enough time step $\Delta t$.
We defer the reader to \cite{palenzuela18} for further details on the numerical schemes and an extensive analysis of the performance with different discretization schemes for different problems, including MHD.

We note that, in previous works, numerical codes designed to handle Hall dynamics show numerical stability issues caused by the the quadratic dispersion relation of the whistler waves present on the system.
Recent work by \cite{gonzalezmorales2018}, for instance, include stabilizing techniques introduced in \cite{osullivan06} for the time advance of the non-linear terms. These techniques, namely the Super Time-Stepping and the Hall Diffusion Schemes, allow to maintain the stability and efficiently speed up the time evolution when the ambipolar or the Hall term dominate and the equations become essentially a parabolic system. Another usual, but less elegant technique in eMHD to control the Hall-generated instabilities is the use of high-order dissipation (also called hyper-resistivity in this context) \citep{huba03,vigano12a}, and/or special recipes for the time advance (e.g., a semi-implicit time advance in the toroidal field component, which acts as an effective high-order dissipative term, \cite{vigano12a}). A similar effect (i.e., a filter of the high-frequency modes which cannot be accurately resolved by our numerical grids) can be achieved by applying artificial Kreiss-Oliger (KO) dissipation to our evolved fields $\{\vec{B}, \Phi\}$ along each coordinate direction~\cite{Calabrese:2004}. A sixth-order derivative dissipation operator, which does not spoil the convergence order of our numerical scheme, can be written in 1D as
\begin{eqnarray}
Q^x_d\, {\bf U}_{i}
&=& \frac{\eta_{ko}}{64 \Delta x}
\bigl( {\bf U}_{i-3}  -6 \, {\bf U}_{i-2} + 15 \, {\bf U}_{i-1}  \nonumber \\
&&-20 \, {\bf U}_{i} 
+ 15\, {\bf U}_{i+1} - 6 \, {\bf U}_{i+2} + {\bf U}_{i+3}  \bigr) ~. \nonumber 
\end{eqnarray}
where $\eta_{ko}$ is a positive, adjustable parameter controlling the amount of dissipation added.\footnote{Note also that FDOC methods can be actually interpreted as a variant of KO with a local (instead of global) dissipation factor.}

The instabilities are especially significant when using $5^{\rm th}$-order methods (i.e., FDOC5, WENO5 and MP5), which need a high dissipation factor to be stabilized, with a potential loss of accuracy. For this reason, in this paper we restrict ourselves only $3^{\rm rd}$ order schemes, that do not require any additional artificial KO dissipation. However, we do not discard using higher-order schemes in the future with an adequate KO dissipation. Unless otherwise indicated, hereafter we will employ WENO3YC~\cite{yamaleev09} without KO dissipation.

In order to test the convergence order for a given solution $u(x,t)$ (where $u$ is a scalar or vector component), we consider the numerical discretized solutions $U_\delta$ obtained with three different resolutions $\Delta x_\delta$, with $\delta=0,1,2$, such that $r=\Delta x_0/\Delta x_1=\Delta x_1/\Delta x_2 > 1$ (typically, $r=2$). The convergence order $n_c$ of the numerical solution at a given time can be defined as
\begin{equation}\label{eq:convergence_def}
  n_c(t) = \log_r \left(\frac{||U_0 - U_1||}{||U_1 - U_2||}\right)
\end{equation}
where $||U_m - U_n|| = \Sigma_{\vec{i}} |U_m^{\vec{i}} - U_n^{\vec{i}}|$ is the L1-norm of the difference of the two discretized fields, and the sum is performed over all indices $i_k$, k=1...N, identifying the N-dimensional grid with the lowest resolution.

\section{Benchmark Tests}\label{sec:tests}

We now discuss the results for a battery of simulations performed to test individual terms of the induction equation, before addressing the general problem. All simulations are run in 2D or 3D Cartesian grid, in rectangular/square or cubic dominion, respectively. The list of tests include those performed in \cite{vigano12a} for a finite-difference 2D axisymmetric code, plus a new one designed to test the behavior of the ambipolar term.

\subsection{Whistler waves}\label{sec:whistler}

\begin{figure}[ht] 
    \centering
    \includegraphics[width=\linewidth]{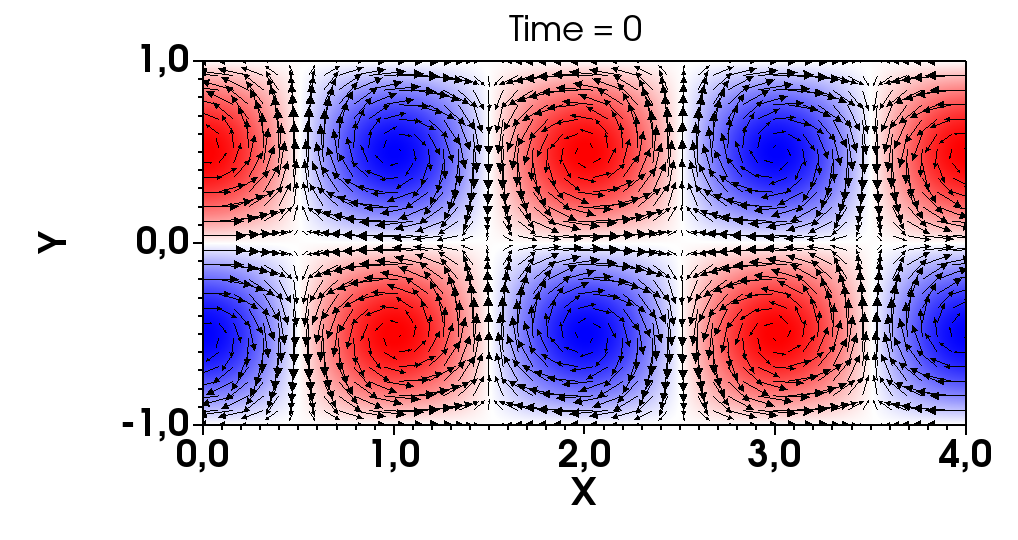}\\ 
    \includegraphics[width=\linewidth]{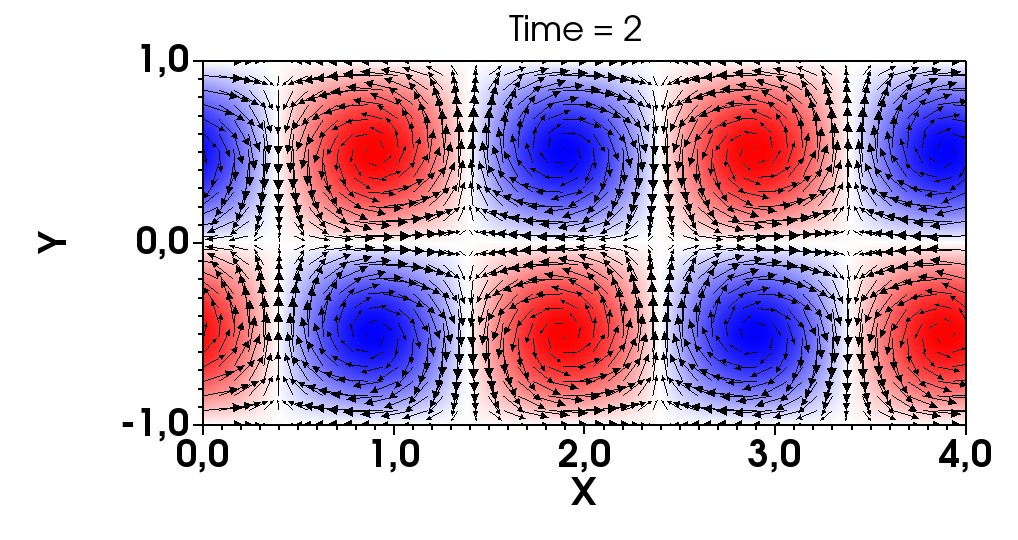}\\ 
    \includegraphics[width=\linewidth]{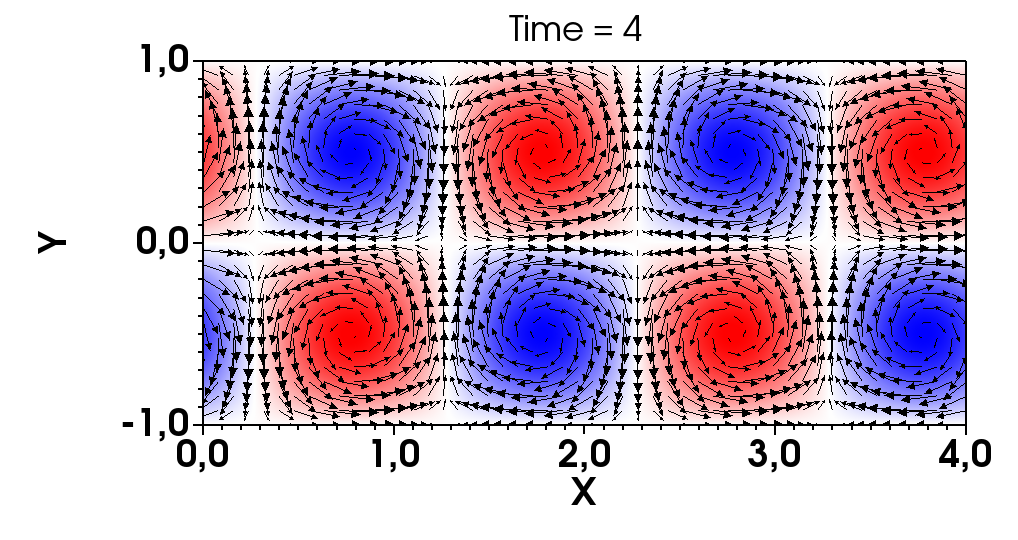}\\ 
    \caption{Whistler wave solution for the case with $256\times 128$ points with the WENO3 scheme, at $t=\{0,2,4\}$ (in units of $\tau_{0}$). WENO3YC, which is our standard choice, gives almost indistinguishable results. The component $B_z$ is shown in blue and red, while black arrows represent the {$(j_{x}, \ j_{y})$} vector field.} 
  \label{fig:whistler} 
\end{figure}

\begin{figure}[ht] 
    \centering
    \includegraphics[width=\linewidth]{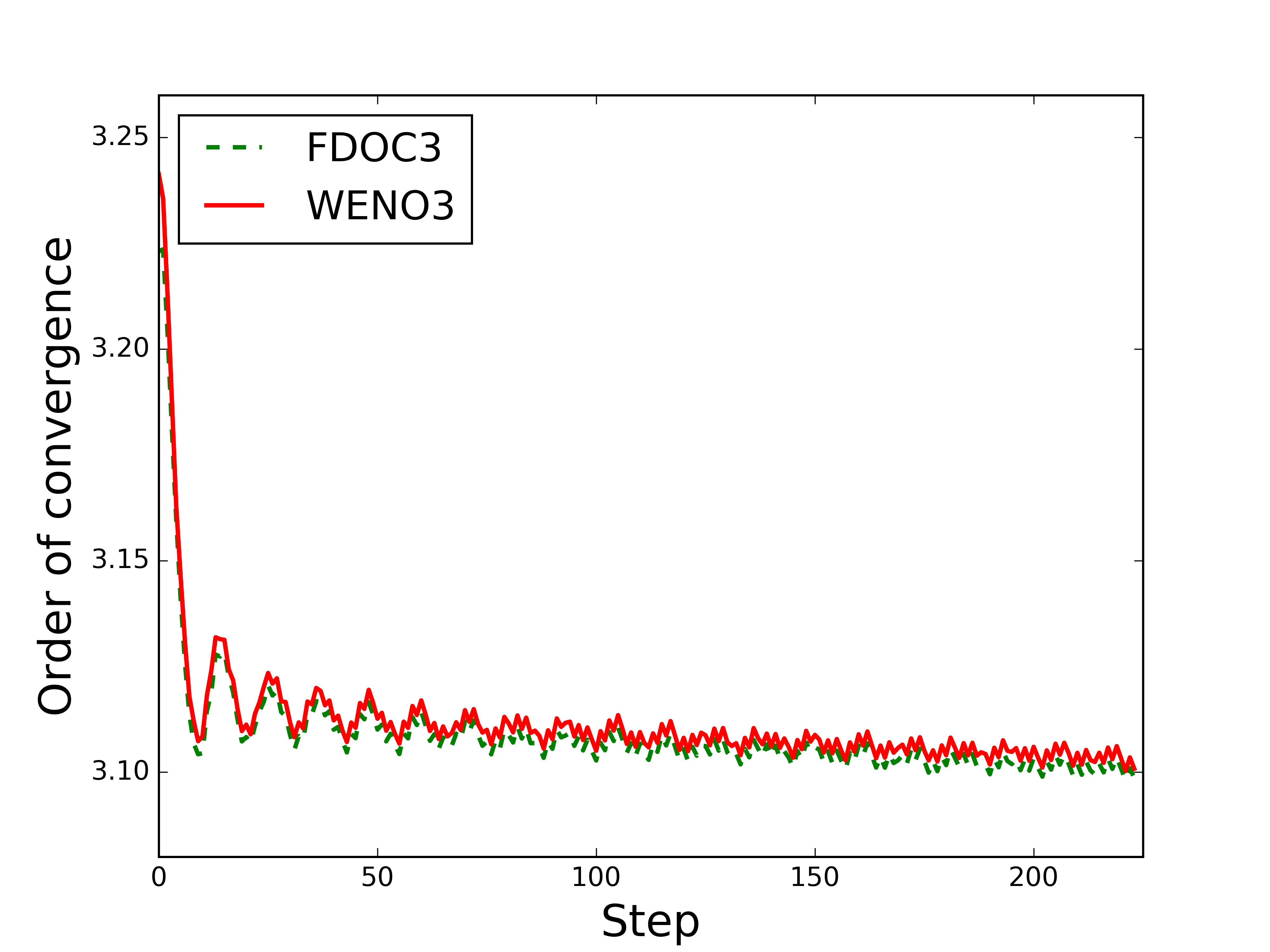} 
    \caption{Whistler wave solution: convergence order for the two methods of 3$^{\rm rd}$-order.} 
  \label{fig:convergence} 
\end{figure}

We first consider a smooth solution for the induction equation with only the Hall term, setting $f_h=1$ and $f_d=f_a=0$. We consider a whistler wave in the linear regime, discussed in \S~\ref{sec:characteristic}, propagating as a perturbation of a homogeneous background magnetic field $\vec{B}_0 = B_0~\hat{x}$. As shown in \ref{app:whistler}, we can find particular circularly polarized solutions with an amplitude $|\vec{B}_1|\ll B_0$, and a generic propagation vector $\vec{k}$, as long as it is not aligned with $\vec{B}_0$. We consider the case $\vec{k}=k~(\hat{x}+\hat{y})/\sqrt{2}$, corresponding to the following initial values of the magnetic field:

\begin{eqnarray}\label{whistler_data}
&& B_x = B_0 + B_1\cos(k y)\cos(k x) \nonumber\\
&& B_y = B_1\sin(k y)\sin(k x) \\
&& B_z = \sqrt{2}B_1\sin(k y)\cos(k x) \nonumber
\end{eqnarray}
As seen above, this solution will propagate along the $x$ direction with velocity
\begin{equation}
 v_w = \sqrt{2}f_h k B_0
\end{equation}
We set up a periodic 2D box $x\in [0,4 L]$, $y\in [-L,L]$, with $L=1$, $B_0=1$, $B_1=10^{-3}$, $k=n\pi/L$, similarly to \cite{vigano12a}. We run a simulation following several crossing times. In Fig.~\ref{fig:whistler} we show for the resolution $(256 \times 128)$, for which we employ a time-step $\Delta t=6.4 \times 10^{-2}$.  the solutions at three different times, given in units of the Hall timescale, which is defined as $\tau_{0}=f_{h} L^{2}/B_{0}$. We have checked that the propagation speed is correct, and the numerical solutions converge to the analytical one for increasing resolutions. Fig.~\ref{fig:convergence} shows that both WENO3 and FDOC3 methods have $3^{\rm rd}$-order convergence, calculated comparing the $B_y$ components in simulations with $(64\times 32)$, $(128 \times 64)$ and $(256 \times 128)$ points. We have checked that, for this particular case, the $\vec{\nabla}\cdot\vec{B}=0$ constraint is maintained numerically at the round-off error without any need for divergence cleaning.

\subsection{Hall drift waves}\label{sec:halldrift}

\begin{figure}[ht] 
    \centering
    \includegraphics[width=\linewidth]{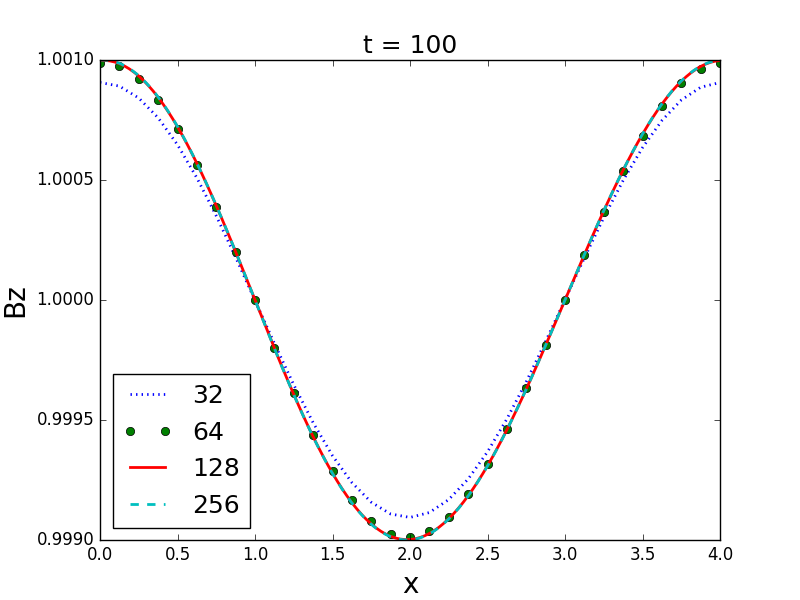} 
    \includegraphics[width=\linewidth]{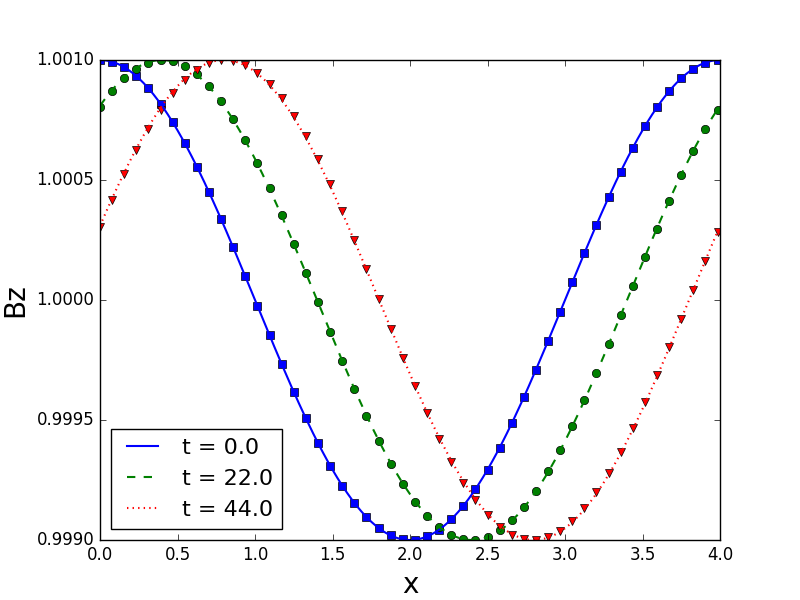} 
    \caption{Hall drift wave solution. Top panel: Comparison of different resolutions with WENO3 at $t=100$. Bottom panel: snapshots at three different times for the resolution 256x128. Lines in the bottom panel represent the analytical solutions while symbols represent the numerical solutions. All times are in units of $\tau_{0}$.} 
  \label{fig:hall_drift} 
\end{figure}

Next, we consider a non-zero vertical gradient of the Hall coefficient, $f_h = f_{h0}(1+\beta y)$, with again $f_d=f_a=0$ in the same 2D rectangular, periodic domain, $x\in [0,4 L]$, $y\in [-L,L]$.
We set the initial values of the magnetic field to:
\begin{eqnarray}
 && B_z = B_0 + B_1\cos(kx) \nonumber\\
 && B_x = B_y = 0~, 
\end{eqnarray}
with $B_0=1$, $B_1=10^{-3}$, $f_{h0}=1$, $\beta=0.2$ and $k=\pi/(2L)$. This solution, as shown in \S~\ref{sec:characteristic} corresponds to a Hall-drift wave propagating in the direction of $\vec{\nabla} f_h \times \vec{B}_0$ (i.e., positive x-direction in our case) with velocity
\begin{equation}
  v_d =\frac{\beta L^{2}}{\tau_{0}} = \beta f_h B_0 = 0.2.
\end{equation}
In Fig.~\ref{fig:hall_drift} we show results using WENO3, with $L=1$. The top panel compares four different resolutions at $t=100 \tau_{0}$, after the wave has crossed the domain 5 times (the domain crossing time is $4L/v_{d}=20 \tau_{0}$). The bottom panel shows the evolution at three different times with a resolution given by $256\times 128$ points, with a time-step $\Delta t=6.4 \times 10^{-2}$.
We checked that the $3^{\rm rd}$-order convergence is maintained. Note that we have kept $B_1$ small to be able to follow several periods in the nearly linear regime, before the non-linear evolution leads to the steepening of the sinusoidal wave and the formation of a discontinuity, intrinsic to the Burgers-like mathematical form of this term. Both in this and in the next test, the $\vec{\nabla}\cdot\vec{B}=0$ constraint is zero by construction.

\subsection{Hall current sheet}\label{sec:hall_nl}

For the third test, also with only the Hall term activated, we adopt the same domain and the same stratification of the Hall coefficient that we have used in the previous test, $f_h(y) = f_{h0}(1+\beta y)$, and the following initial conditions:
\begin{eqnarray}
 && B_z= B_0 \cos(k x) \nonumber\\
 && B_x = B_y = 0
\end{eqnarray}
with $k=\pi/L$. In this case, we can write the induction equation as a Burgers equation for $B_z$: $\partial_t B_z = \beta B_z \partial_x B_z$ (see the detailed analysis in \cite{vigano12a}). Any smooth solution will form discontinuities. For our initial conditions, the expected breaking time (at which the solution reaches its maximum steepness before starting dissipating) is $\tau_b = \lambda/(4\beta) = 2.5$, where $\lambda = 2\pi/k = 2$ is the wavelength of the solution.

\begin{figure}[ht] 
    \centering
    \includegraphics[width=\linewidth]{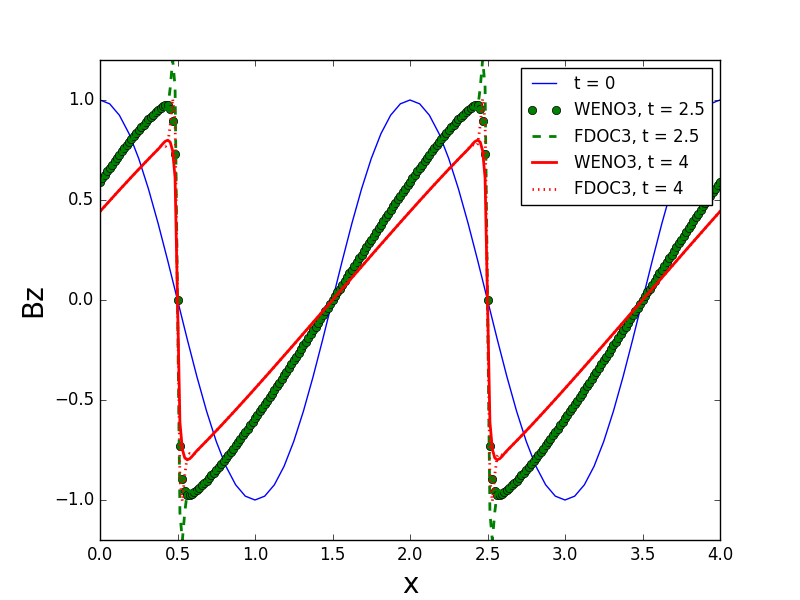} 
    \includegraphics[width=\linewidth]{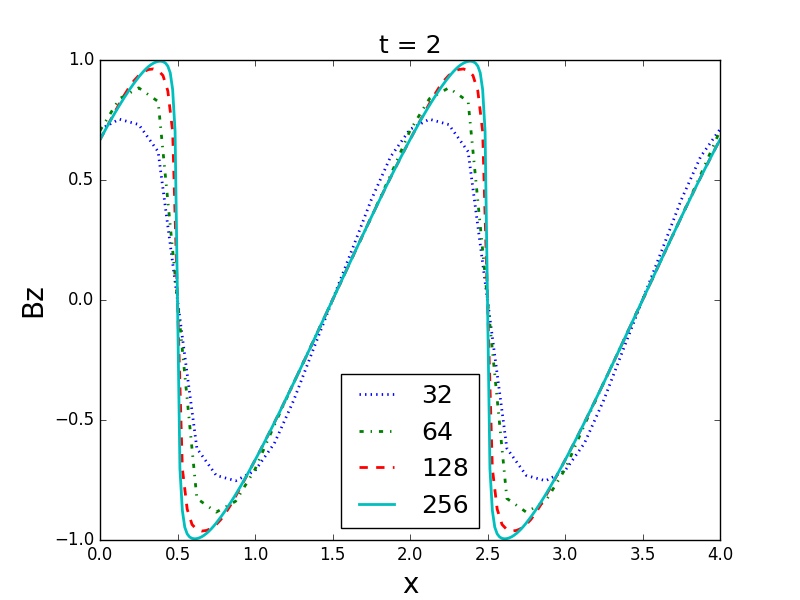} 
    \caption{Hall current sheet solution. Comparison of the behavior for FDOC3 and WENO3 at different times (top), and solution at $t=2$ for different resolutions with WENO3 (bottom). Times are given in units of $\tau_{0}$. Note that the spikes in the FDOC3 case correspond to a single-point deviation from the solution.} 
  \label{fig:hall_shock} 
\end{figure}
This test is more challenging, because of the full non-linearity and the formation of a current sheet  after the initial sinusoidal profile steepens. In Fig.~\ref{fig:hall_shock} we compare the different resolutions (different $N$ shown) and we note that FDOC3 is able to follow the test, but it does not accurately reproduce the shock profile, due to the presence of a single-point spike. WENO3, instead, follows it, even though some numerical dissipation eventually damps the shock. Of course, the presence of a shock implies a downgrading of the order of convergence. The breaking time in the simulations converges toward the theoretical one for increasing resolutions. For $t=2$ (shown in the bottom panel of Fig.~\ref{fig:hall_shock}), the low-$N$ cases have already formed and started to dissipate the shock due to the lack of spatial resolution.

\subsection{Ohmic dissipation}\label{sec:bessel}

We turn now to the study of purely resistive modes, by setting $f_d=1$, $f_a=f_h=0$. Simulations are run in a cubic 3D Cartesian domain $[-1,1]^3$. We consider the particular case of the self-similar solution, eq.~(\ref{eq:selfsimilar_bessel}). In axial symmetry, the resistive eigenfunctions are provided by the spherical Bessel functions of index $l$ (indicating the multipole). The initial values for the dipolar $l=1$ case in spherical coordinates are:
\begin{eqnarray}\label{eq:bessel}
  && B_r=\frac{B_0}{\xi^2} \left(\frac{\sin \xi}{\xi}-\cos \xi\right)\cos\theta~,\nonumber\\
  && B_\theta=\frac{B_0}{2 \xi^2}\left(\frac{\sin{\xi}}{\xi}-\cos \xi-\xi\sin{\xi}\right)\sin\theta~,\\
  && B_\varphi=\frac{B_0 }{2 \xi} \mu\left(\frac{\sin \xi}{\xi}-\cos \xi\right)\sin\theta~,  \nonumber
\end{eqnarray}
where $\xi=\mu r$, $B_0$ is a normalization factor, and the free parameter $\mu$ controls the amount of twist of the solution.  We map this initial model to our Cartesian grid by the standard transformation, $r=\sqrt{x^2+y^2+z^2}$, $\theta=\cos^{-1}(z/r)$, $\varphi=\tan^{-1}(y/x)$, with the azimuthal direction contained in the $(x,y)$ plane:
\begin{eqnarray}
 && B_x = B_r \sin\theta\cos\varphi + B_\theta\cos\theta\cos\varphi - B_\varphi\sin\varphi  \nonumber \\
 && B_y = B_r \sin\theta\sin\varphi + B_\theta\cos\theta\sin\varphi + B_\varphi\cos\varphi  \nonumber\\
 && B_z = B_r \cos\theta - B_\theta\sin\theta  \label{eq:sphtocart}
\end{eqnarray}
In the limit $\xi =\mu r \rightarrow 0$, we recover the solution corresponding to a homogeneous vertical field $\vec{B} \rightarrow (B_0/3) \hat{z}$. 

\begin{figure}[ht] 
    \centering
    \includegraphics[width=\linewidth]{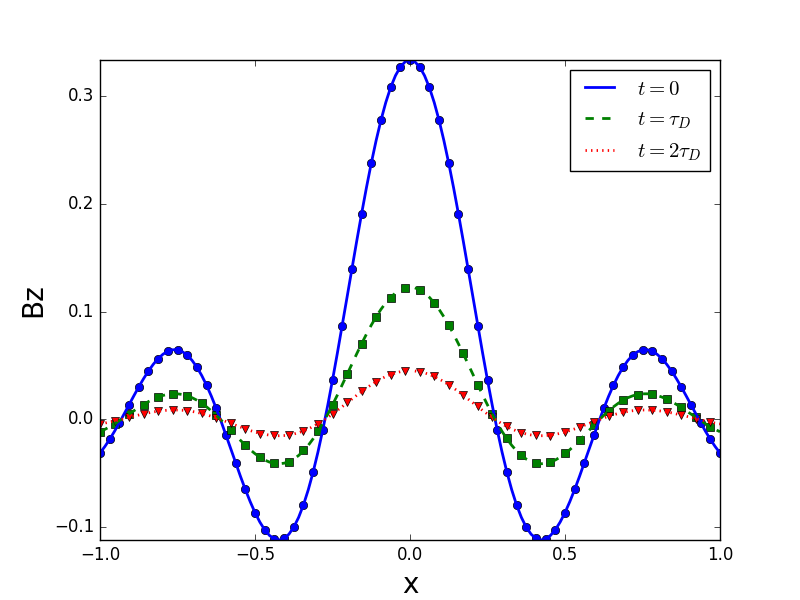} 
    \includegraphics[width=\linewidth]{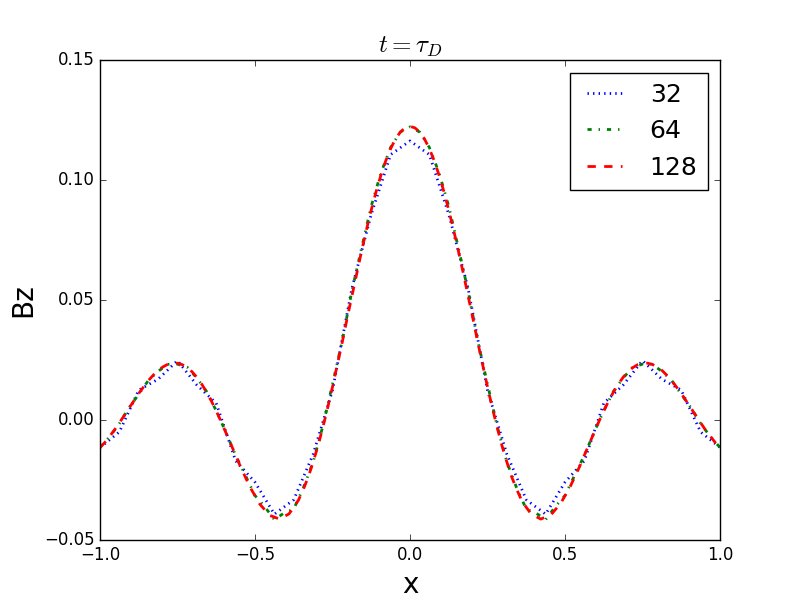} 
    \caption{Purely resistive self-similar solutions with WENO3 in a central section plane. Evolution of the solution for $N=64^3$ (top), and solution at $t=\tau_d$ for different resolutions (bottom). Lines in the top panel represent the analytical solutions while symbols represent the numerical solutions.}
  \label{fig:bessel} 
\end{figure}

At the boundaries, we impose the $\exp(-t/\tau_d)$ analytical solution of eq.~\ref{eq:selfsimilar_bessel}. We follow the evolution of the modes on several Ohmic timescales, until the field has decreased to a few orders of magnitude below the initial values. In Fig.~\ref{fig:bessel}, we show the evolution of the $B_0=1$, $\mu=1$ case for $N=64^3$ at three different times (top), and the solution at $t=\tau_d$ for different resolutions (bottom). Despite the relatively low resolution, the errors inherent to mapping solutions in spherical coordinates to Cartesian grids, we find an excellent agreement with the analytical solution. We also found that the convergence order is about two for $3^{\rm rd}$-order and $5^{\rm th}$-order schemes here tested. The difference with the nominal convergence reached in the smooth cases above, according to our interpretation, lies on the numerical errors introduced in the discretized version of current. While for a smooth, small perturbation (whistler, Hall drift waves), the leading error is the spatial discretization, for a general case, where there is no background magnetic field, the source of error is likely to come from the calculation of the current. In this respect, see \ref{app:j} for a discussion of different discretized expressions of current. In this case, the divergence of magnetic field tends to slowly grow, so that the divergence cleaning needs to be activated to achieve the maximum accuracy (see \ref{sec:poloidal} for a discussion of the divergence cleaning parameters).

\subsection{Ambipolar diffusion. The Barenblatt-Pattle solution}\label{sec:ambipolar_test}

\begin{figure}[ht] 
    \centering
    \includegraphics[width=\linewidth]{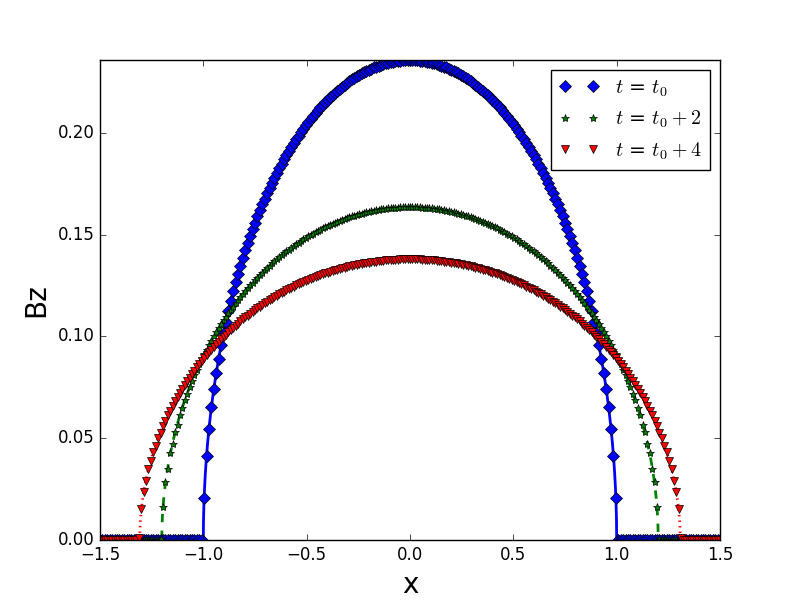} 
    \includegraphics[width=\linewidth]{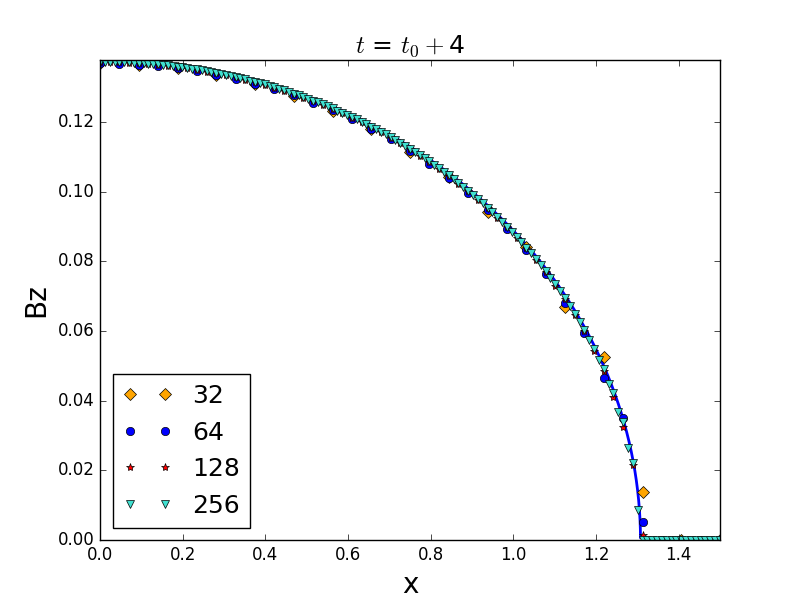} 
    \caption{Barenplatt-Pattle solution for different times for $N=256^2$ (top), and for different resolutions at $t=t_{0}+4$ (bottom). The apparent tail for the low-resolution cases is a visual effect, due to the connection of the points across the front. Lines correspond to the analytical solutions and symbols represent the numerical solutions.} 
  \label{fig:ambipolar} 
\end{figure}

To test the ambipolar term, we now set  $f_a$ to a constant value, and $f_d=f_h=0$. We use a particular analytical solution which is axially symmetric in cylindrical coordinates with only one component of the magnetic field depending on the radial distance only ($B_z(\varpi)$, where $\varpi^2=x^2+y^2$), so that initially the currents are perpendicular to the magnetic field, so that $(\vec{j}\times\vec{B})\times\vec{B}=-B^2\vec{j}$, and its evolution can be expressed as

\begin{eqnarray}
  \derparn{t}{B_z} && = -f_a\{\vec{\nabla}\times[B_z^2 (\vec{\nabla}\times B_z \hat{z})]\}\cdot\hat{z}  \nonumber \\
   && = f_a \frac{1}{\varpi}\frac{\partial}{\partial\varpi}\left[ \varpi\left(B_z^2 \frac{\partial B_z}{\partial\varpi}\right) \right]
\end{eqnarray}
This form is analogous to a non-linear diffusion equation 
\begin{equation}
u_t = \vec{\nabla}\cdot(mu^{m-1}\vec{\nabla} u)~,
\end{equation}
where $m$ is a power index. The analytical, 2D solutions proposed by Barenblatt and Pattle \citep{barenblatt52,pattle59} consist in a delta function of integral $\Gamma$ at the origin, which diffuses outwards with finite velocity (the diffusion front is clearly defined, contrarily to the infinite-speed of the parabolic standard diffusion equation). The analytic solution can be explicitly written as follows:

\begin{equation}
   u(\varpi,t)=\max\left\{0, t^{-\alpha}\left[\Gamma - \frac{\alpha(m-1)}{2dm}\frac{\varpi^2}{t^{\frac{2\alpha}{m}}}\right]^{\frac{1}{m-1}}\right\}
\end{equation}
where $d$ is the dimension of the problem, and $\alpha=(m-1+2/d)^{-1}$. The initial pulse spreads with a front located at

\begin{equation}
\varpi_f(t)=\left(\frac{2\Gamma dm}{\alpha(m-1)}\right)~t^{\alpha/d} 
\end{equation}
In our case, we test a solution for $B_z$ in the $xy$-plane (an expanding cylindrical flux tube), for which $d=2$, $m=3$, $\alpha=1/3$, $f_a=3$ so that the evolution is analytically given by:

\begin{equation}
 B_z(x,y,t) = t^{-(1/3)}\left[\Gamma - \frac{1}{18}\frac{(x^2+y^2)}{t^{1/3}}\right]^{1/2}~.
\end{equation}
We set $t_0=1$ as the beginning of our simulation, and $\Gamma=1/18$, so that the front propagates according to $(x^2+y^2)=t^{1/6}$ (being $x^2+y^2 = 1$ at $t_0=1$). In Fig.~\ref{fig:ambipolar} we show that our results correctly reproduce the analytical values, both in the shape of the spreading pulse and in the correct propagation speed of the front. The sharp discontinuity in the slope of $B_z$ is well-reproduced even for low resolutions. The convergence order in this case is also around 2, as for the resistive case.

\section{Star-like dynamics}\label{sec:3d}

We now consider geometric configurations qualitatively similar to a neutron star crust. We work in a cubic domain $[-L,L]^3$ and we set up a shell, defined by $R_c \leq r \leq R_\star$, where $r$ is the distance from the center of the dominion. We set $R_c=8$ (the core/crust interface) and $R_\star=10$ (the star's surface). This setup will allow to qualitatively compare the results provided by our 3D Cartesian code with the well known Hall dynamics for existing axisymmetric models in 2D spherical coordinates \citep{pons07b,pons09,vigano12a,vigano12b,vigano13,gourgouliatos13,gourgouliatos14a,gourgouliatos14b}.  Note that in a realistic set up, the values of $R_c$ and $R_\star$ depend on both the equation of state and the mass of the star (the larger the mass, the thinner the shell). The thickness of the crust can vary within the range $\sim 1-2$ km, while $R_\star \sim 10-14$ km for a $M=1.4~M_\odot$ star. 

In the shell, we assume $f_a=f_d=0$, and 

\begin{equation}\label{eq:fhall_3d}
 f_{h}(r)=[1-(r-\delta r)^2/R_\star^2]^{-1}~,
\end{equation}
where $\delta r > 0$ is a parameter that controls the steepness of the gradient $df_h/dr$ in the outer crust. The smaller the value of $\delta r$, the larger the gradient. We set $\delta r=0.05$, which gives values of $f_h \in [1,100]$.

Note that, in a realistic neutron star, the typical velocities related to the long-term evolution of the magnetic field in the crust are of the order of km~yr$^{-1}$, while in the magnetosphere the dynamical timescale is given by the speed of light (10-15 orders of magnitude faster than the interior). This transition occurs in a thin (100 m) liquid layer called the envelope. This is the reason why, in realistic simulations, one cannot numerically simulate both the short-time dynamics and the long-term evolution. When the latter is the focus of the study,  one has to place the outer boundary at densities $\rho\sim 10^{10}$ g cm$^{-3}$, and provide appropriate boundary conditions connecting the interior field to some equilibrium state outside (usually, a vacuum or a force-free magnetosphere).

\begin{figure*}[ht]
	\centering
	\includegraphics[width=.4\linewidth]{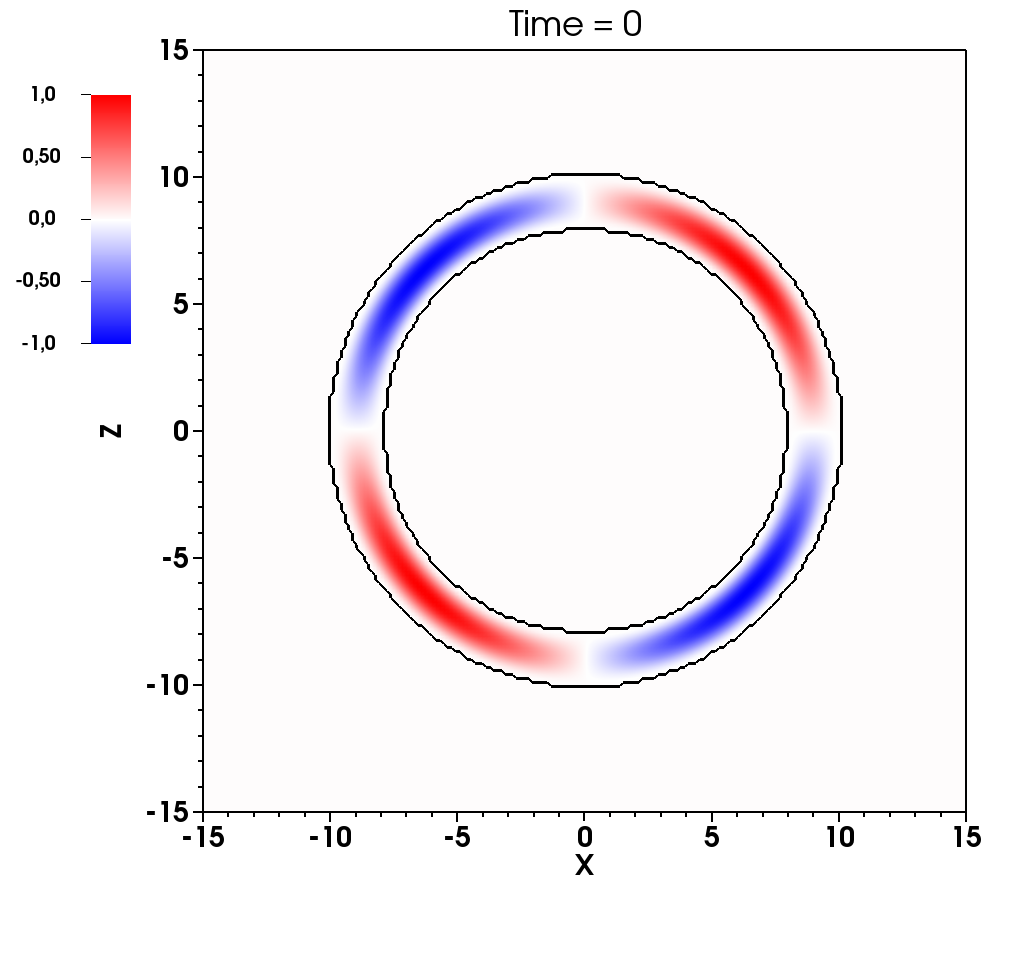} 
	\includegraphics[width=.4\linewidth]{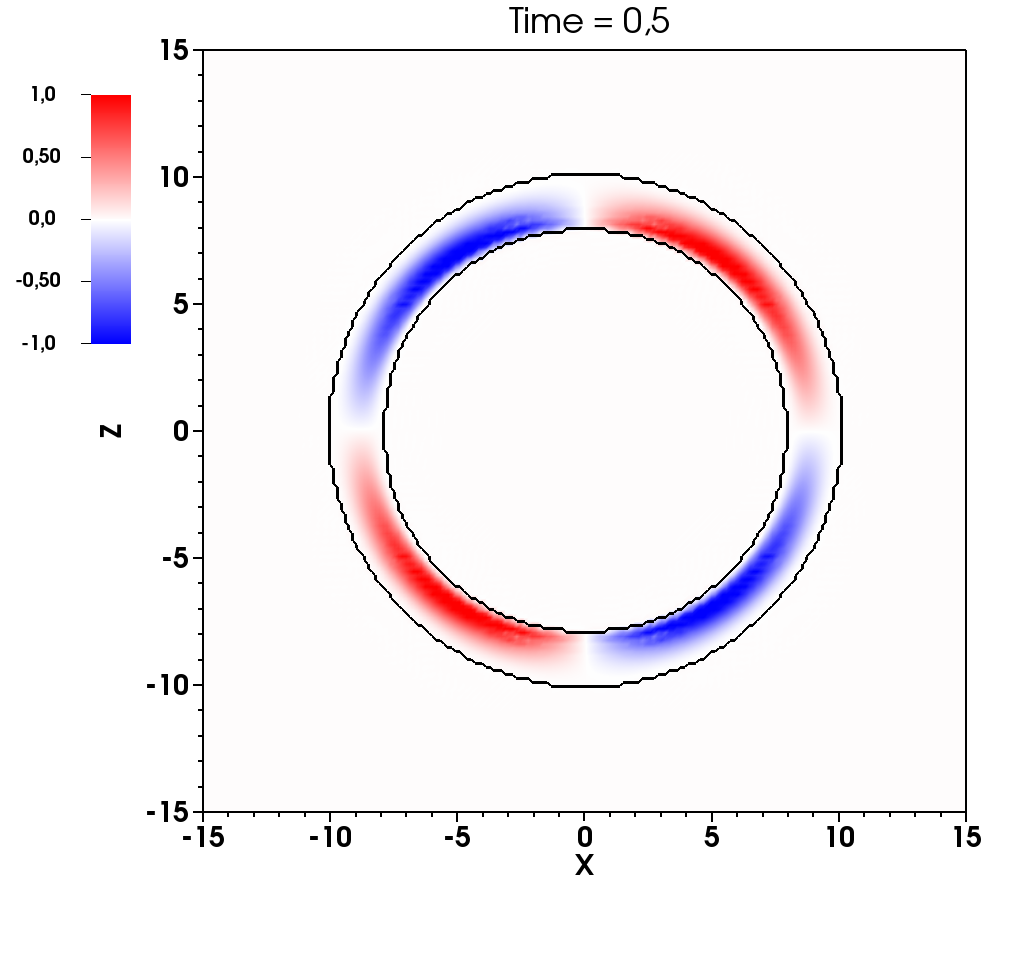}
	\includegraphics[width=.4\linewidth]{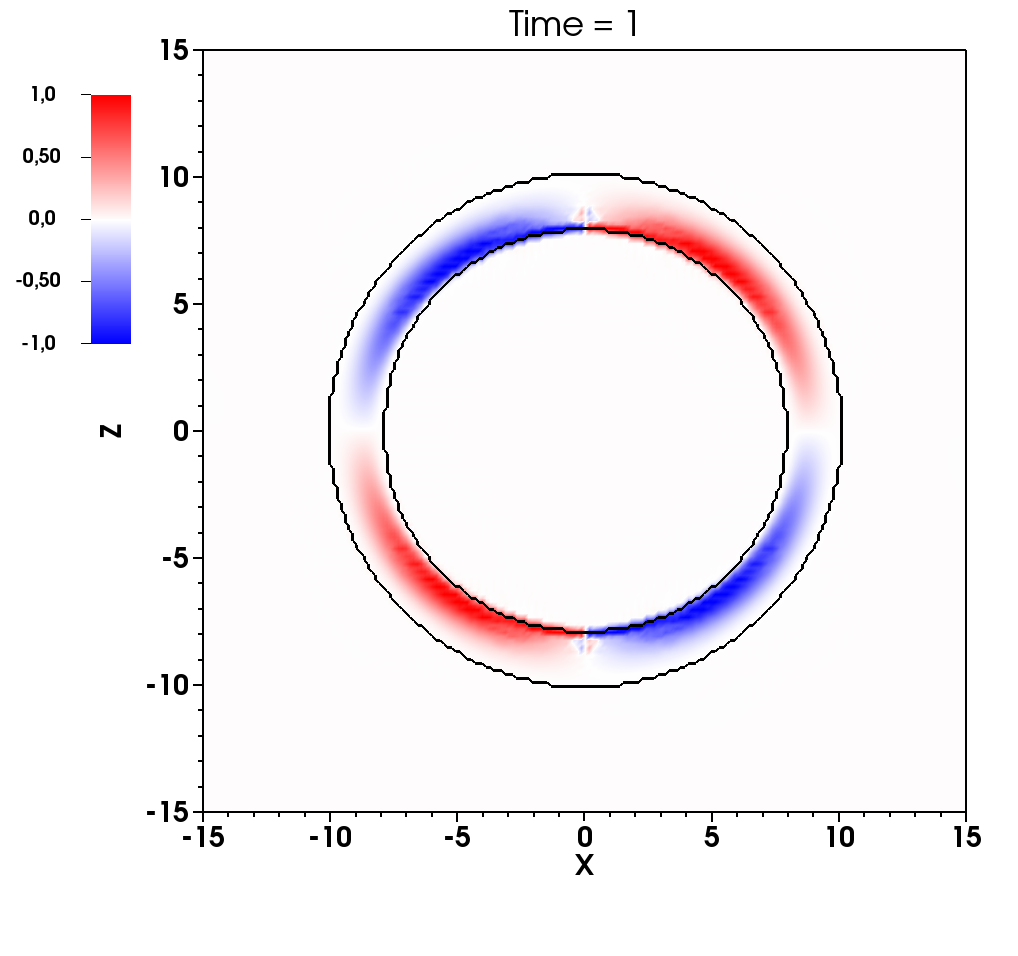}
	\includegraphics[width=.4\linewidth]{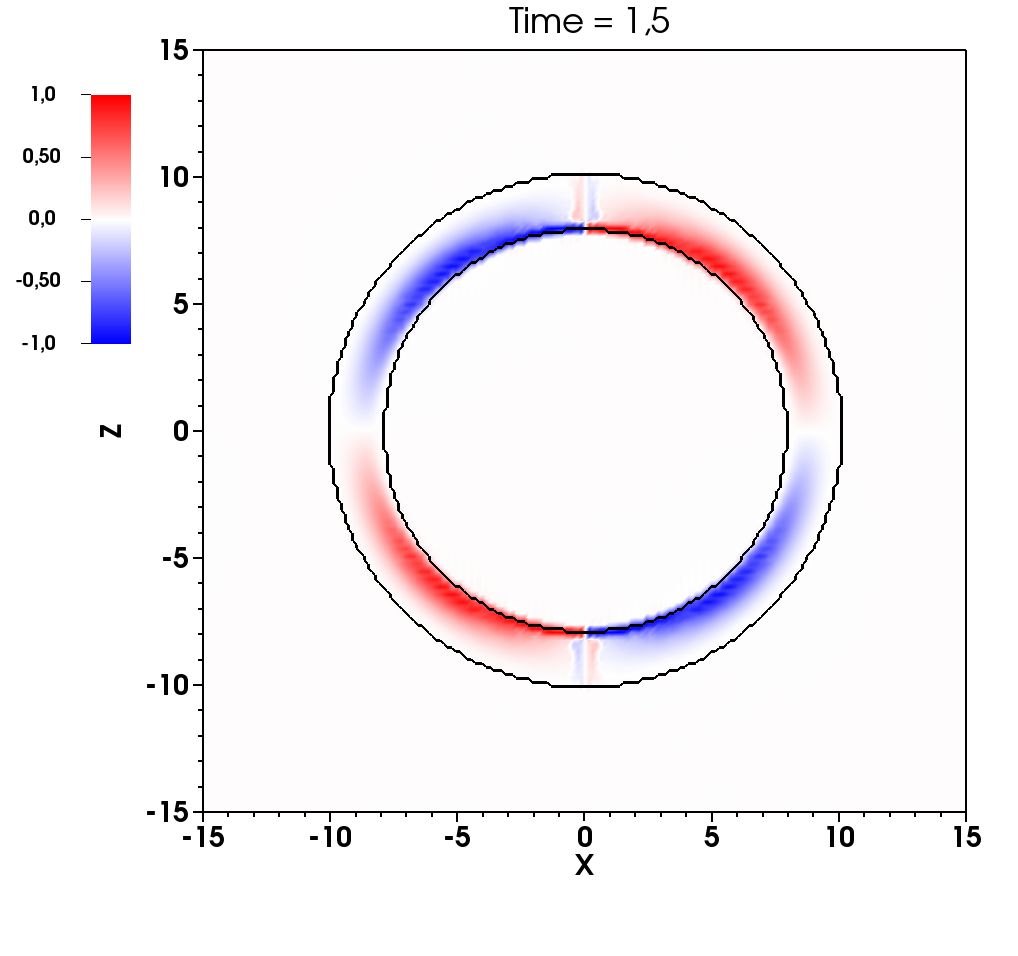}
	\caption{Evolution of the toroidal field (color scale) in the meridional slice (i.e., $B_y$ in the plane ${x-z}$) at different times $t=\{0,0.5,1,1.5\}$ (from top left to bottom right). The black lines represent the boundaries of the crust.}
	\label{fig:toroidal} 
\end{figure*}

We explore two simple cases for the initial magnetic field: either a purely toroidal field, or a purely poloidal field, both confined in the spherical shell and axially symmetric around the $z$-axis. These configurations are not realistic and are prone to different instabilities. As a matter of fact, the magnetic field configuration, likely relaxed to a MHD-equilibrium state, has to be constituted by an axisymmetric mixed poloidal-toroidal configuration, or, more realistically, by a full 3D solution, possibly containing small-scale structures. However, in this paper we study them separately, since each one is a useful test that allows us to compare them with known previous simulations and to underline some important features.

We test both cases with a uniform grid and with a simple mesh refinement (MR2 hereafter), set up as one additional layer with a refinement factor of two (defined as the ratio of the coarse grid sizes to the refined grid size). In a realistic case, we will use many layers, in order to cover a domain with $L\gg R_\star$, and to have a fair resolution in the most dynamical region. Note also that below, for these simple tests, we employ a resolution of about $\Delta x = L/N \simeq 0.15$ km. Note that the typical resolutions of the 2D spherical coordinates production runs of \cite{vigano12a,vigano13} were given by $\sim 60$ angular points (arcs of $\sim 0.5$ km), and $\sim 60$ radial points, which were equally spaced in $\log(\rho)$, meaning that the radial size of the cells varies in the range $\sim 10-100$ m (and larger in the core). Our 3D Cartesian code can reach much better resolutions in all directions, given the adaptability offered by the AMR and the efficiency of the code. Increasing the resolution
is challenging due to the non-linear Hall dynamics, which triggers short-scale, very fast whistler waves, as we discuss below. This is the reason why we typically have to use, for these particular tests, a time-step of $\Delta t \lesssim 10^{-3} \Delta x_c $, where $\Delta x_c = 2L/N$ is the spatial size of the coarse grid.

\subsection{Evolution of a purely toroidal field.}\label{sec:toroidal}

The evolution of a purely toroidal field subject to the Hall drift has been studied previously, like for example in Section 5 of \cite{vigano12a}. We successfully reproduced a few standard cases. For conciseness, here we only show the case of an initial quadrupolar field, with the opposite polarity of the example in \cite{vigano12a}, in order to explore the behavior of the Cartesian grid when the field becomes stronger near the magnetic axis.

We set $L=15$ and an initial toroidal magnetic field given by:
\begin{eqnarray}\label{eq:initial_toroidal}
  B_\varphi = B_{0}\frac{(r-R_c)^2(r-R_{\star})^2}{r} \cos\theta\sin\theta~,
\end{eqnarray}
with the usual spherical to Cartesian transformation, so that $B_x=-B_\varphi\sin\varphi$, $B_y= B_\varphi\cos\varphi$, and $B_z=0$. Outside the shell, we impose $\vec{B}=0$. We use a number of points $N=200$ in each direction, corresponding to $\sim 15$ points covering the thickness of the shell, and we set $c_h=4$ (see next test for a discussion about this parameter).

In Fig.~\ref{fig:toroidal} we display the initial magnetic configuration and its evolution at different times, showing the toroidal component in a meridional slice (i.e., $B_y$ in the $x-z$ plane). The expected evolution consists in a vertical displacement of both toroidal rings toward the poles, in opposite directions. If $f_h$ is constant, the drift is purely vertical and the toroidal field would be displaced until encountering the star's surface, where the current forms a screening sheet at the surface or it propagates outside, depending on the treatment at surface (see next section). However, in presence of a radial gradient of $f_h$ (as in this case), the drift velocity acquires an additional negative radial component toward the star center. Therefore, the evolution proceeds such that the toroidal rings initially drift toward the poles up to a latitude for which the different terms in the velocity drift balance, reaching a sort of equilibrium solution. Since in this case in the core there is no evolution of the field ($f_a=f_h=f_d=0$ for $r<R_c$), a radial discontinuity of the toroidal field forms at the boundary between the crust and the core near the poles, sustained by a screening current flowing in the meridional direction over the $r=R_c$ surface. Most magnetic energy is confined at low magnetic co-latitude, where a stronger ring-shaped magnetic field forms. The magnetic field is zero in the axis, where a radial current forms. We have checked that, throughout the evolution, the integrated constraint ${\cal C}$ is kept at a level below $10^{-5}$ of the integrated magnetic energy ${\cal E}_m$.

Last, we remind that the evolution of a purely toroidal, axisymmetric magnetic field only under the Hall term (i.e., with $f_a=f_d=0$), notoriously implies that the poloidal field is kept identically zero (the opposite is not true). In this respect, note that the use of the Cartesian grid implies that the code does not deal with spherical components and the star's surface is not exactly spherical. These features introduce a spurious poloidal field when one recovers the spherical component from the computed Cartesian ones. We have checked that, with this resolution, the poloidal field is kept at a level $\lesssim 1\%$, and the larger deviations appear at the crust-core interface, where there is a discontinuity of the magnetic field and where also $\vec{\nabla}\cdot\vec{B}$ shows its maximum. However, these deviations do not cause any numerical or physical instability and they are cured by increasing the resolution, which improves the accuracy of the Cartesian discretization of the spherical surfaces.

\subsection{Initially purely poloidal field}\label{sec:poloidal}

\begin{figure}[ht] 
    \centering
    \includegraphics[width=\linewidth]{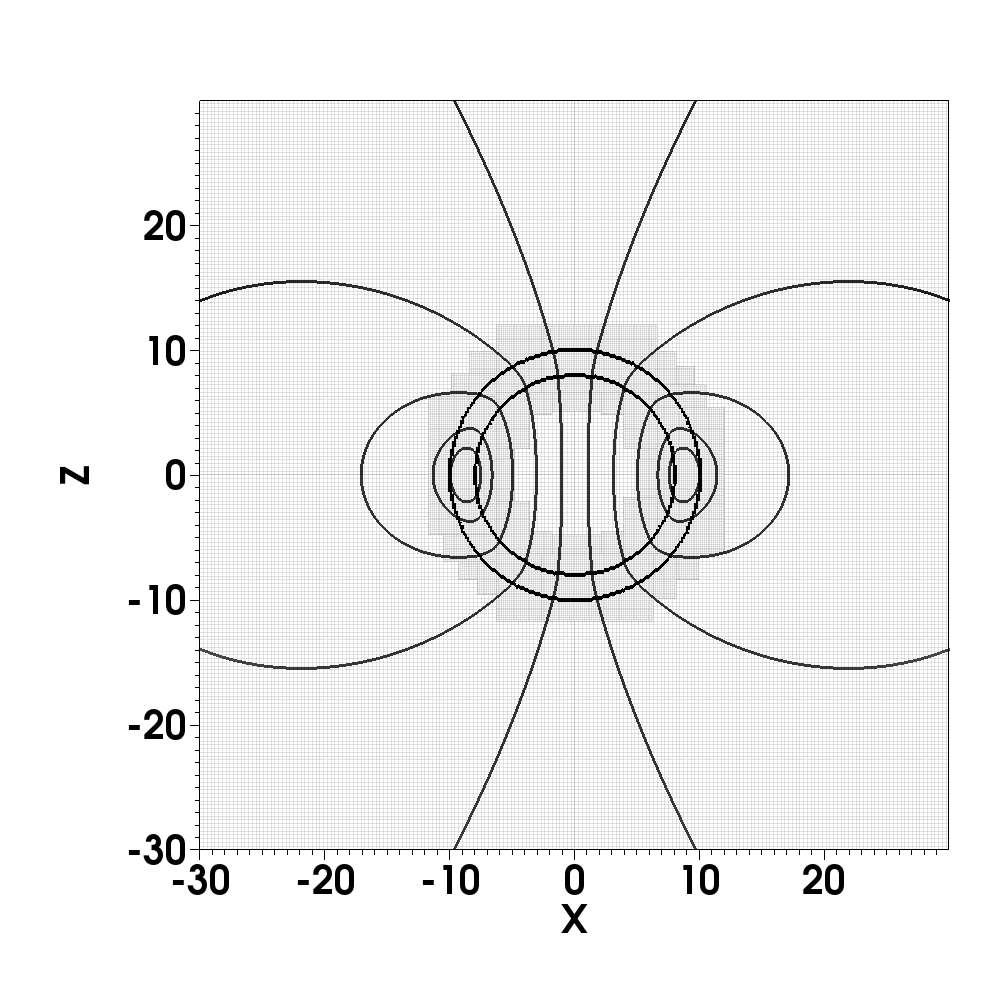} 
    \caption{Purely poloidal field: initial data in the plane perpendicular to the $y$-axis. Black lines indicate the poloidal components of the field, $B_x-B_z$. The position of the shell is indicated by the thick line. The numerical mesh is also shown in gray, with the MR2 in the crustal region.} 
  \label{fig:poloidal_initial} 
\end{figure}

\begin{figure*}[ht] 
    \centering
    \includegraphics[width=.32\linewidth]{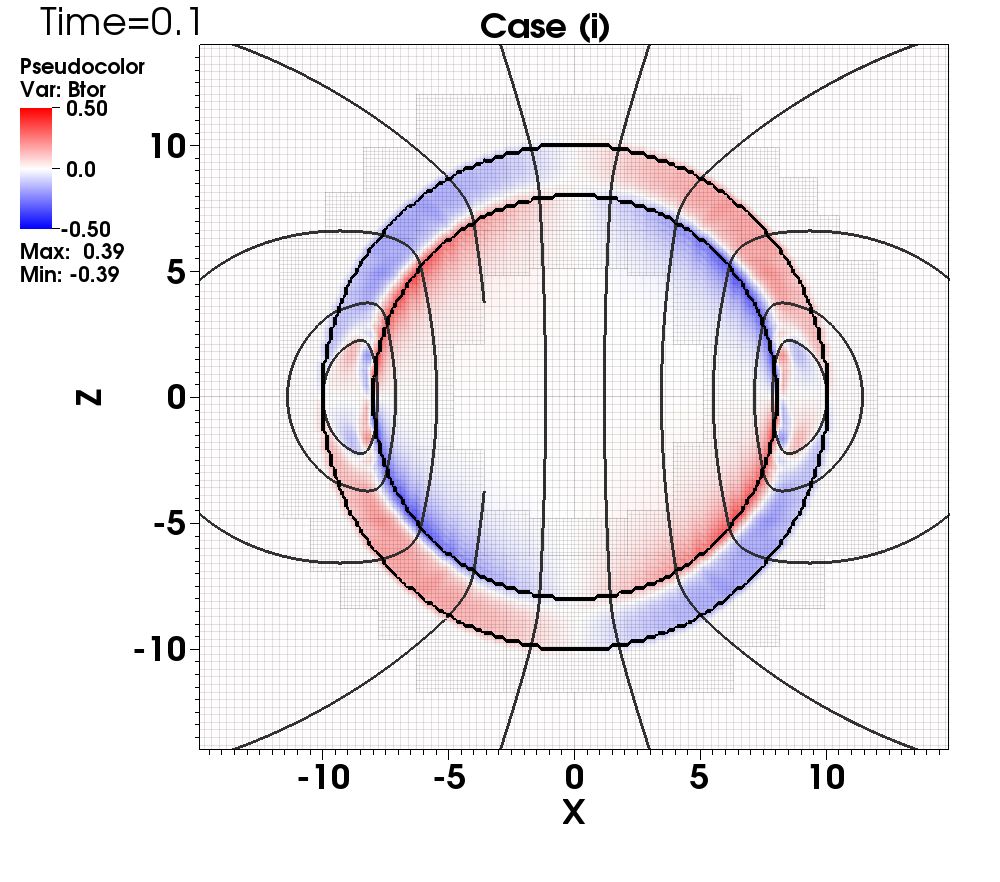}
    \includegraphics[width=.32\linewidth]{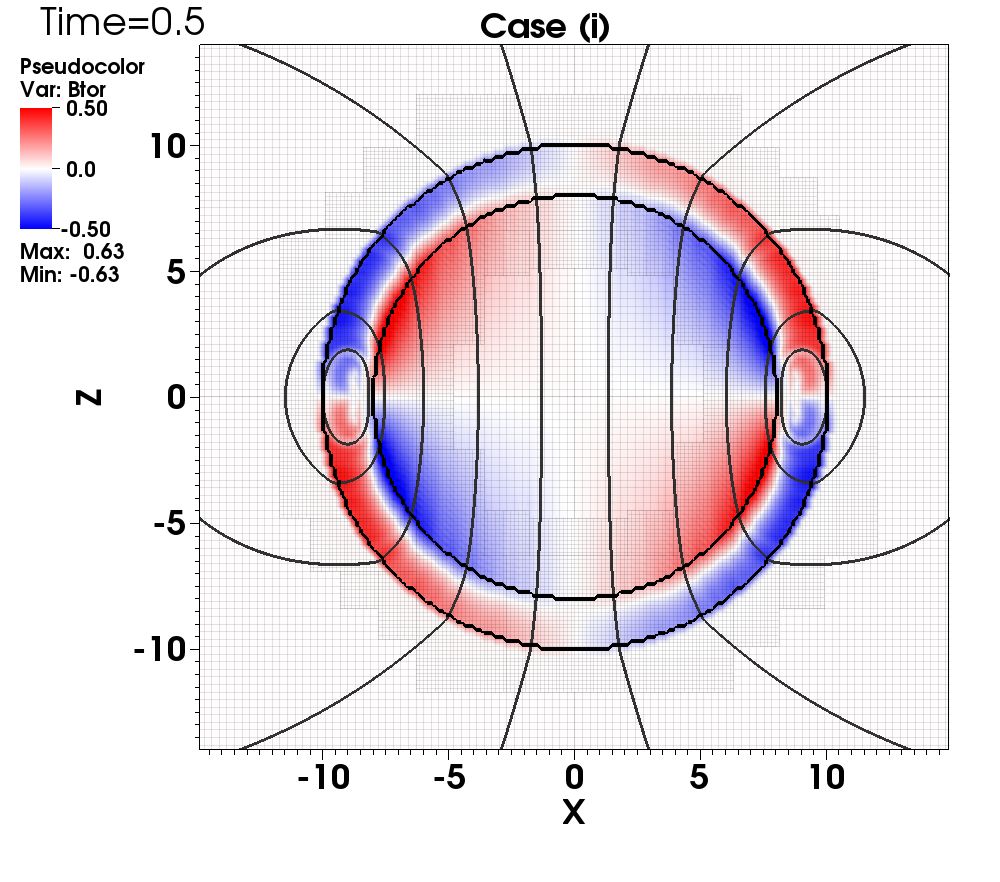}
    \includegraphics[width=.32\linewidth]{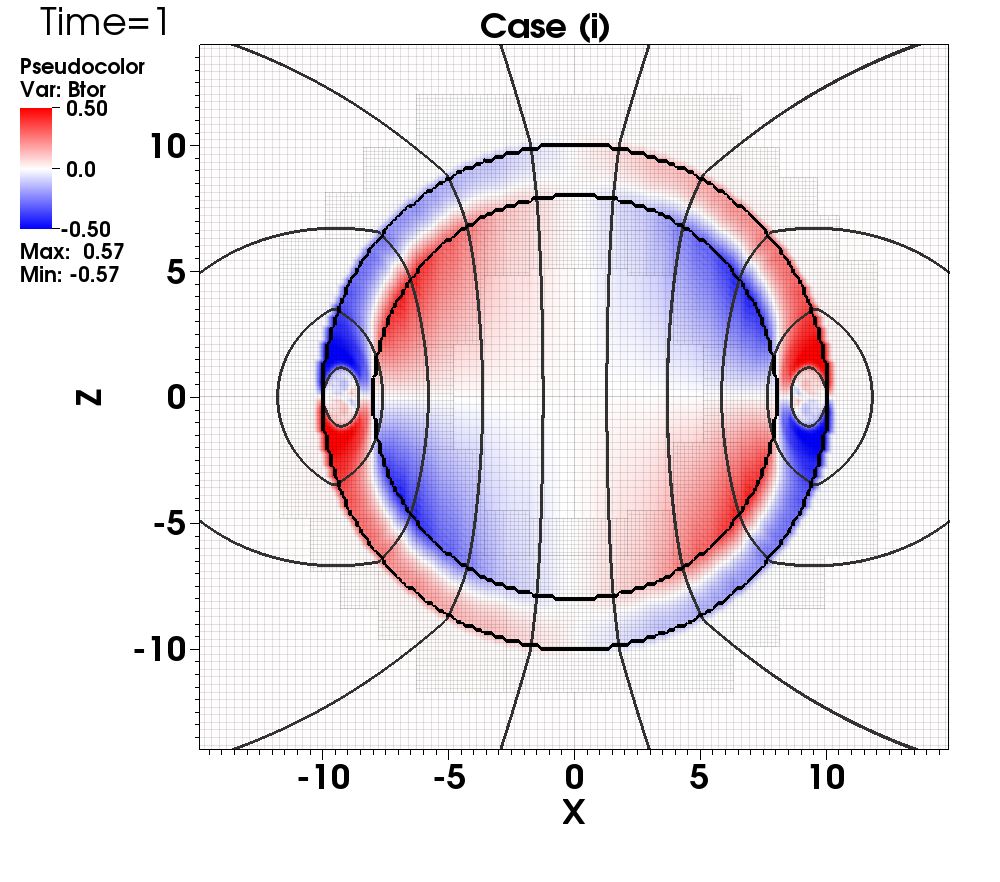}\\
     \includegraphics[width=.32\linewidth]{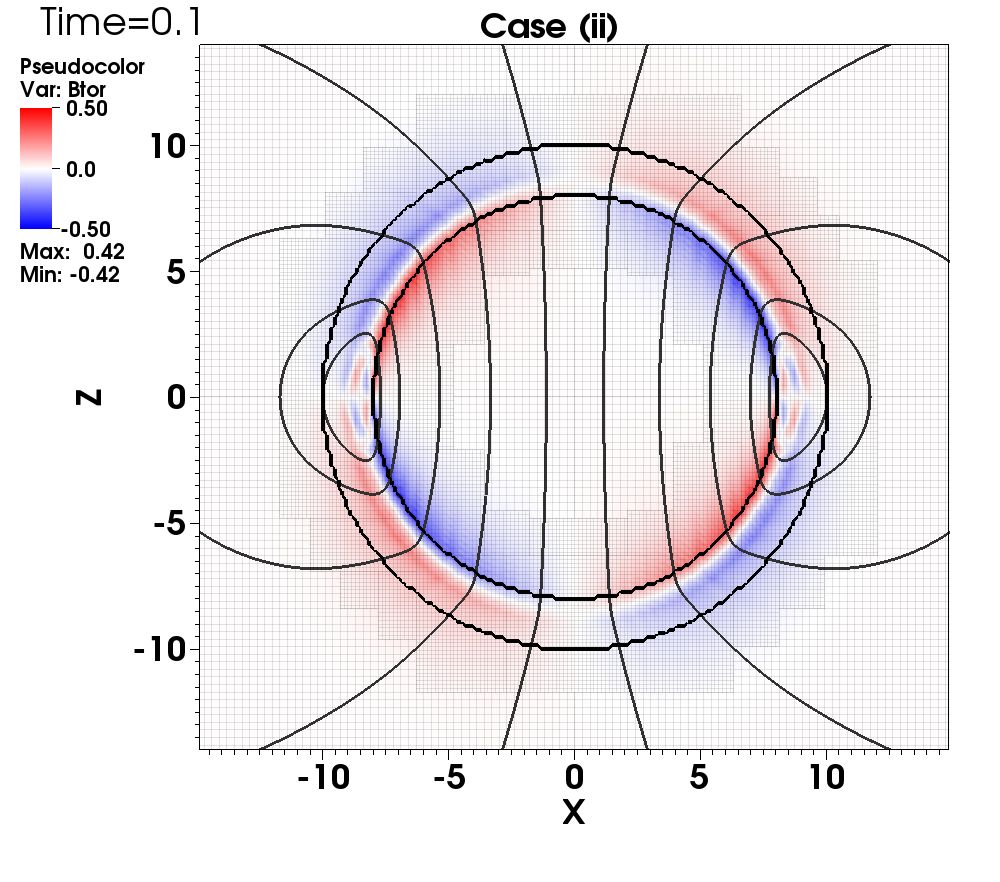}
     \includegraphics[width=.32\linewidth]{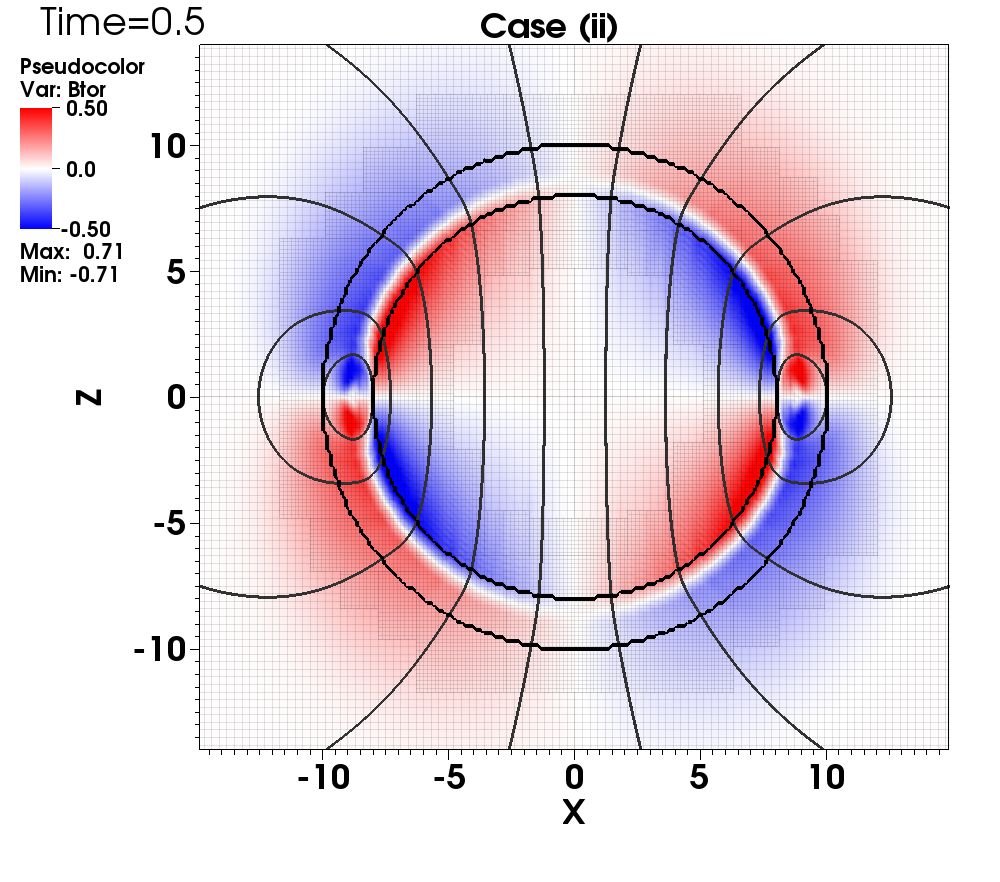}
     \includegraphics[width=.32\linewidth]{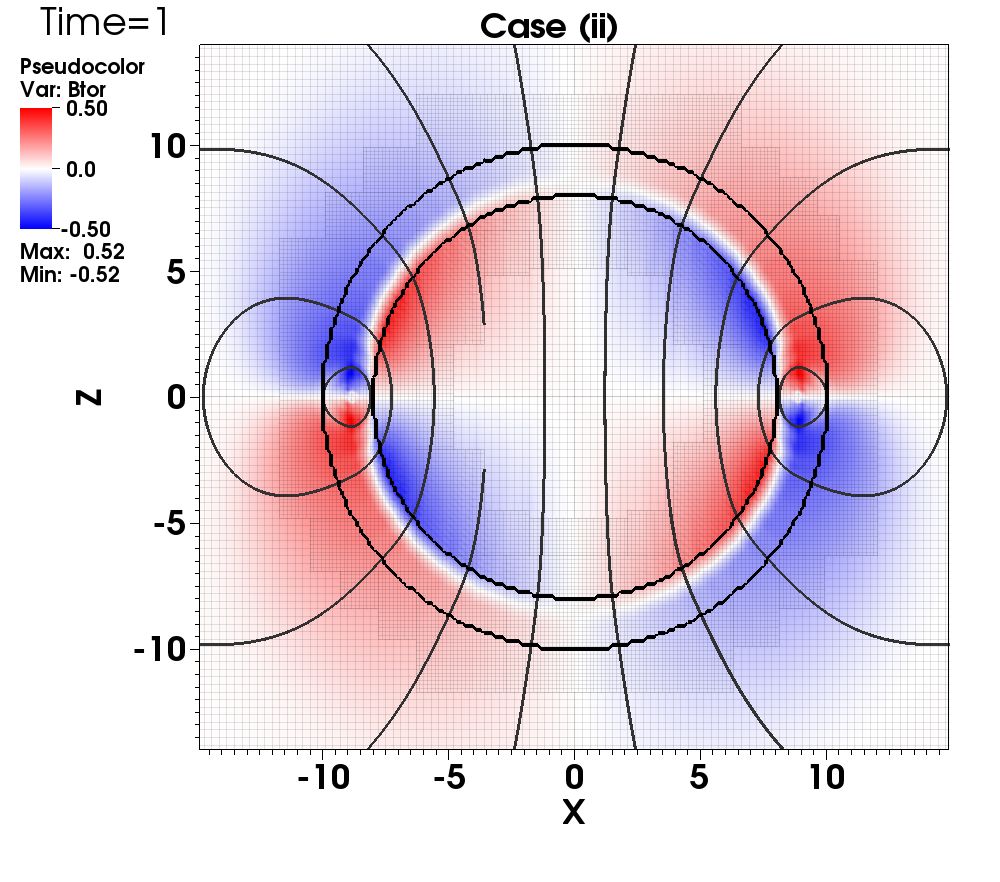}
    \caption{Evolution at $t=0.1$ (left), $t=0.5$ (center), $t=1$ (right) of the initially purely poloidal field case, for case (i) (top panels) and case (ii) (bottom). We show the plane perpendicular to the $y$-axis, the poloidal field, the shell boundaries and the numerical mesh are indicated as in Fig.\ref{fig:poloidal_initial}. The color scale indicates the toroidal field ($B_y$ component). Note the kinks in the magnetic field lines at the star's surface in case (i).} 
  \label{fig:poloidal_evo} 
\end{figure*}

\begin{figure*}[ht] 
    \centering
    \includegraphics[width=.32\linewidth]{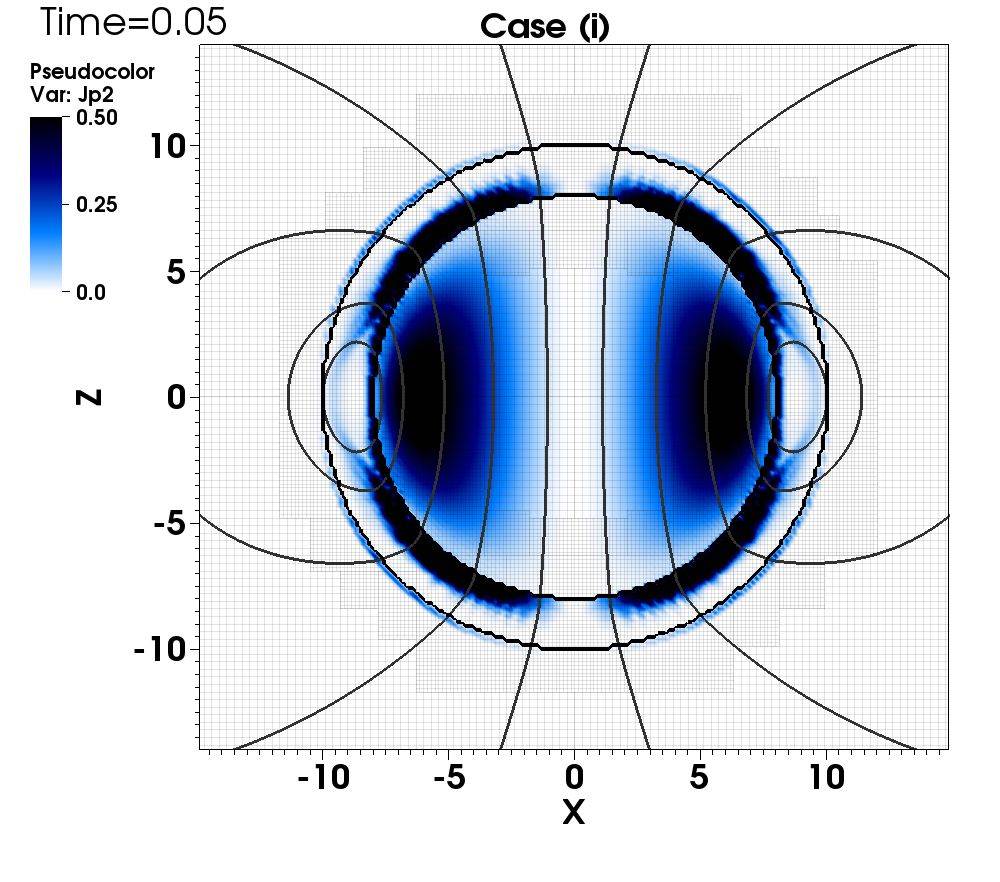}
    \includegraphics[width=.32\linewidth]{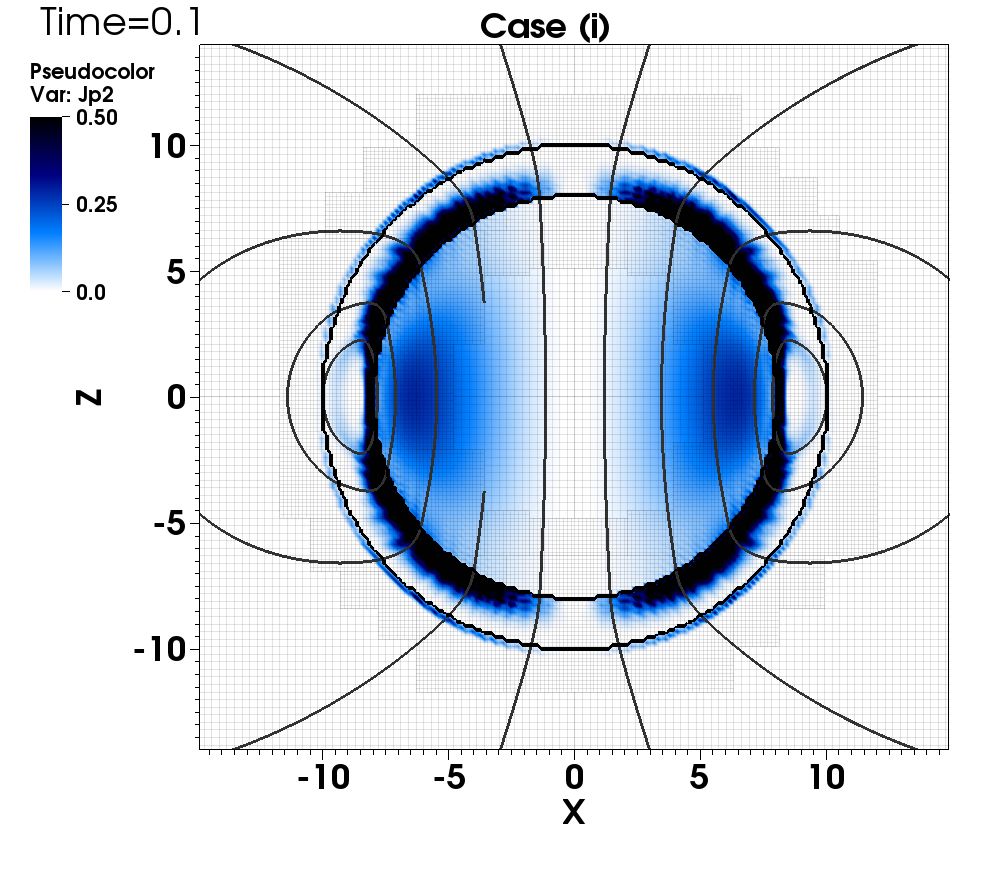}
    \includegraphics[width=.32\linewidth]{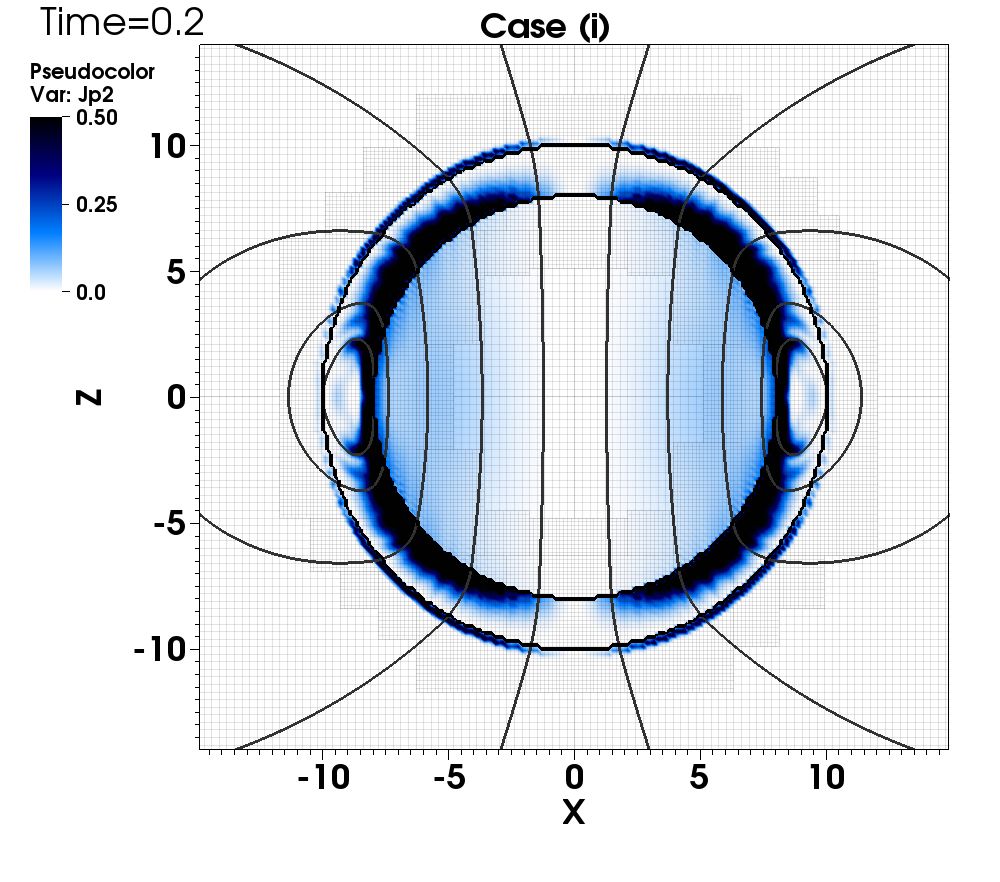}\\
    \includegraphics[width=.32\linewidth]{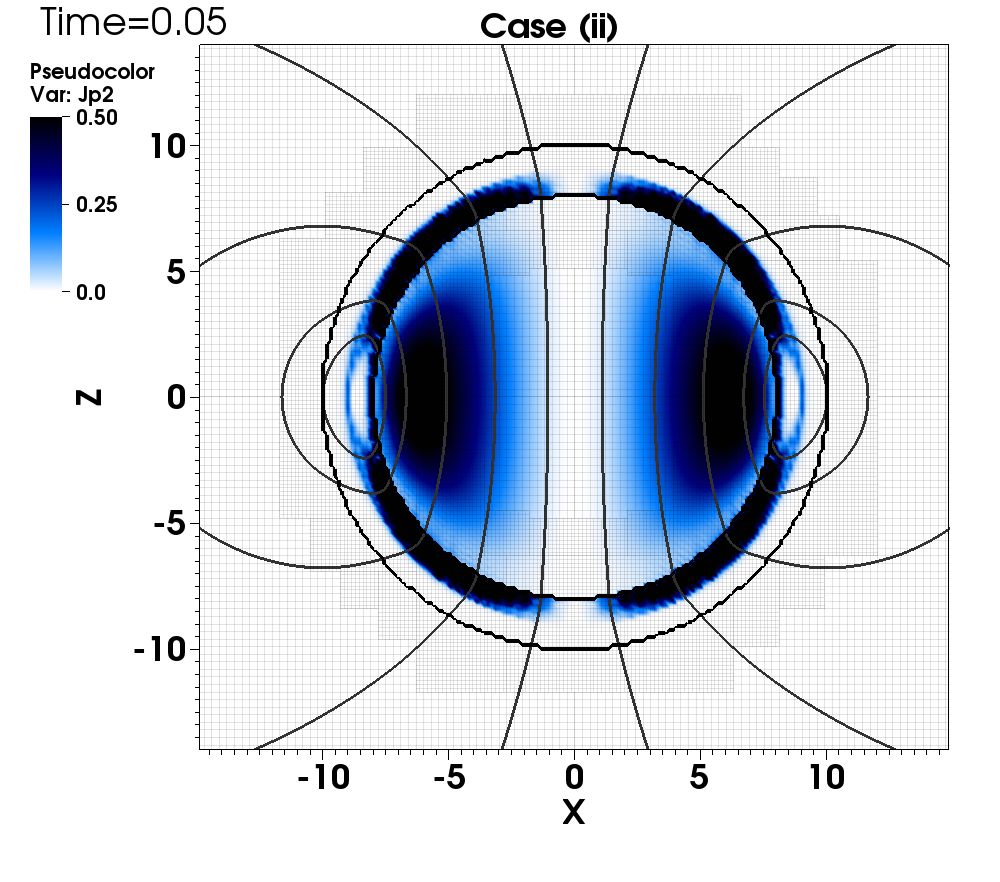}
    \includegraphics[width=.32\linewidth]{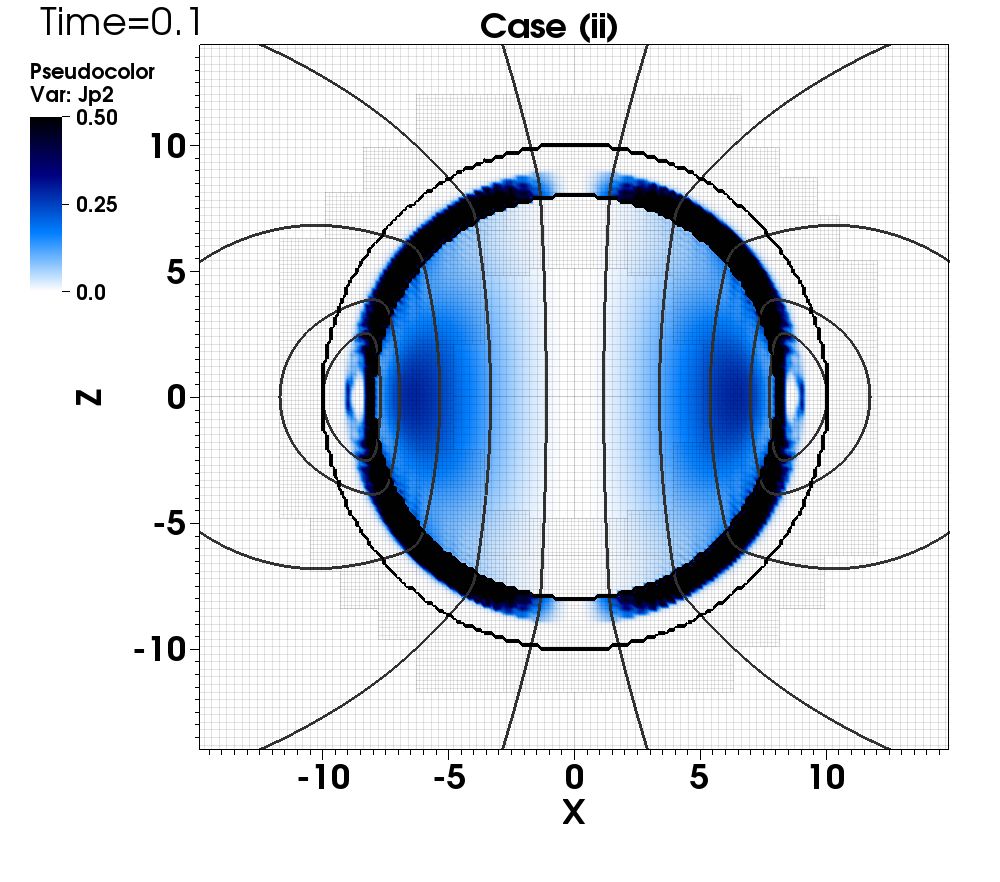}
    \includegraphics[width=.32\linewidth]{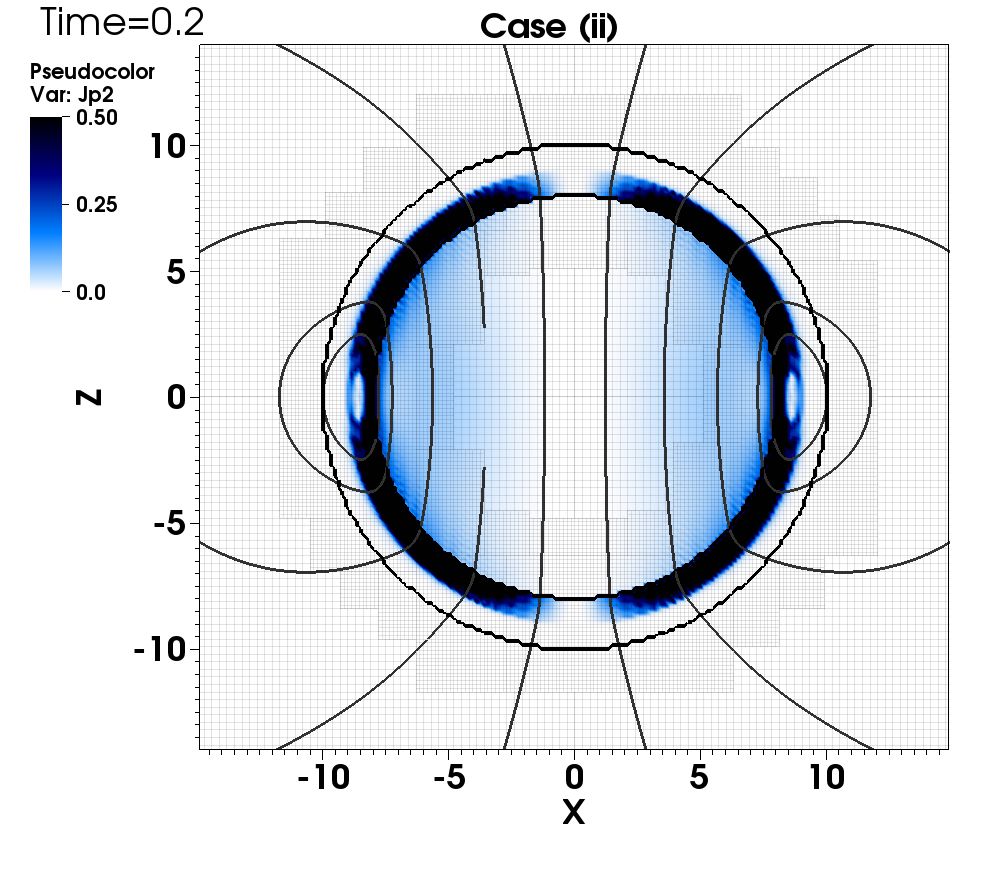}
    \caption{The same as Fig.~\ref{fig:poloidal_evo}, but with he white-blue-black scale indicating the squared perpendicular current, $j_\perp^2 = |\vec{j}-(\vec{j}\cdot\vec{B})\vec{B}/B^2|^2$, at $t=0.05$ (left), $t=0.1$ (center), $t=0.2$ (right). Note the formation of the strong current sheet on the star's surface in case (i).} 
  \label{fig:poloidal_jperp_evo} 
\end{figure*}

\begin{figure}
    \centering
    \includegraphics[width=.8\linewidth]{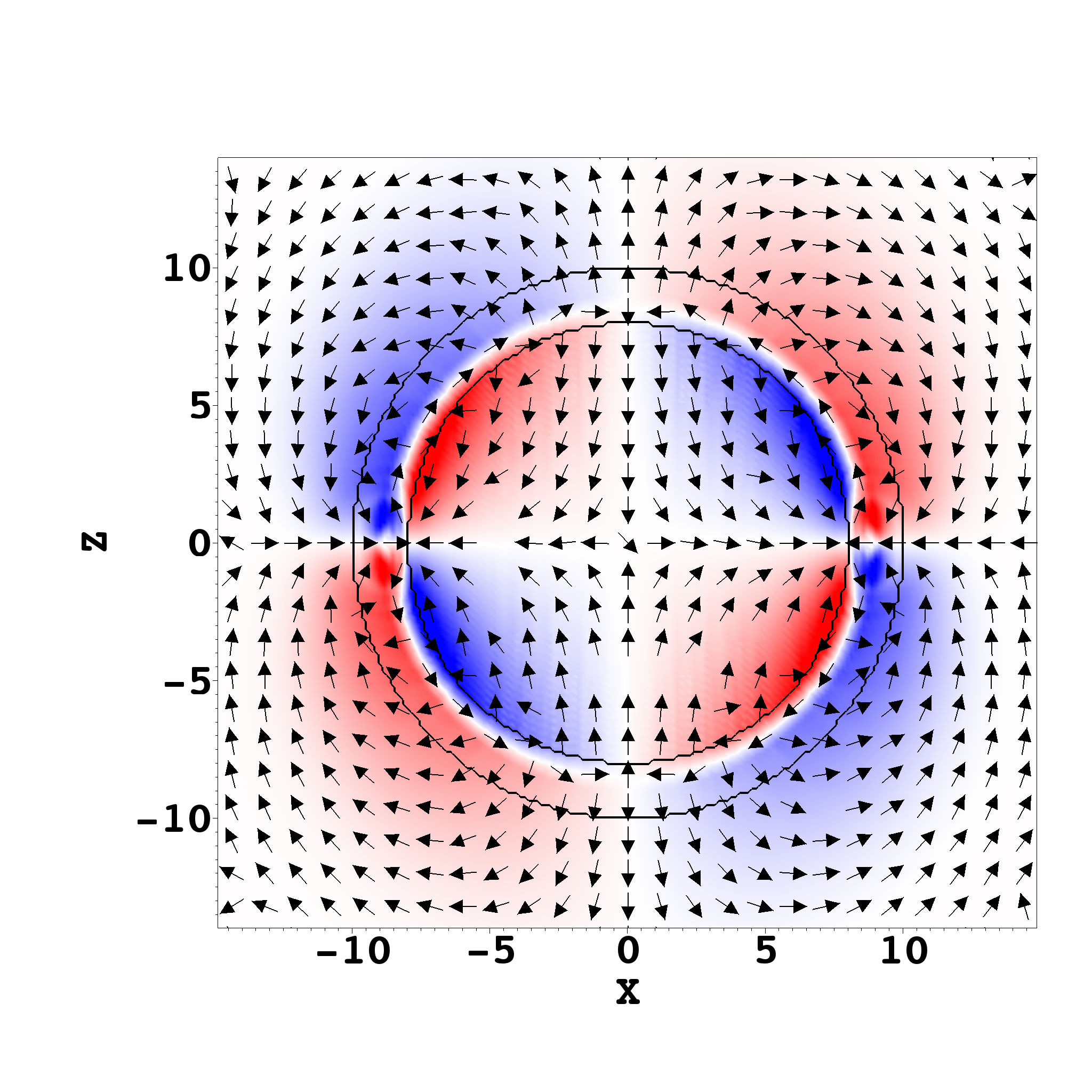}
    \caption{System of currents (black arrows, representing the local direction) and toroidal field (color scale as in Fig.~\ref{fig:poloidal_evo}) for the case (ii) at $t=0.5$.}
  \label{fig:poloidal_currents} 
\end{figure}

\begin{figure*}[ht] 
	\centering
	\vspace{0.5cm}
	\includegraphics[width=.32\linewidth]{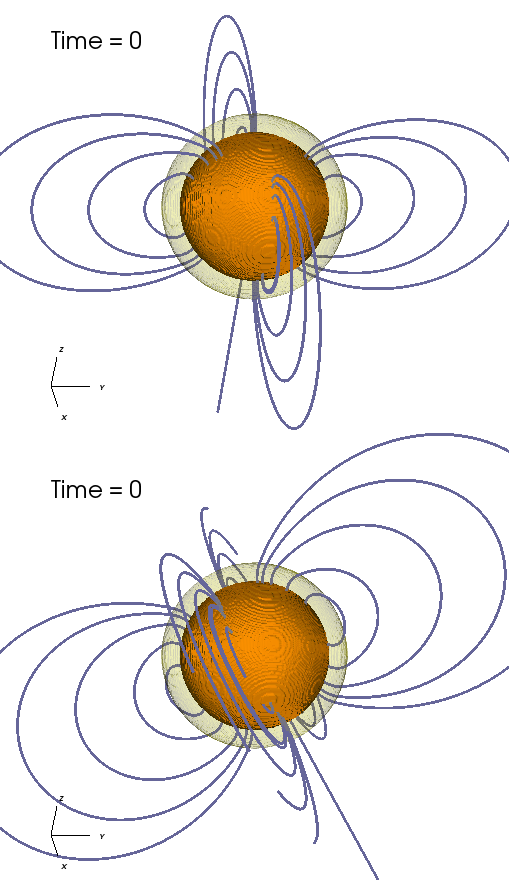} 
	\includegraphics[width=.32\linewidth]{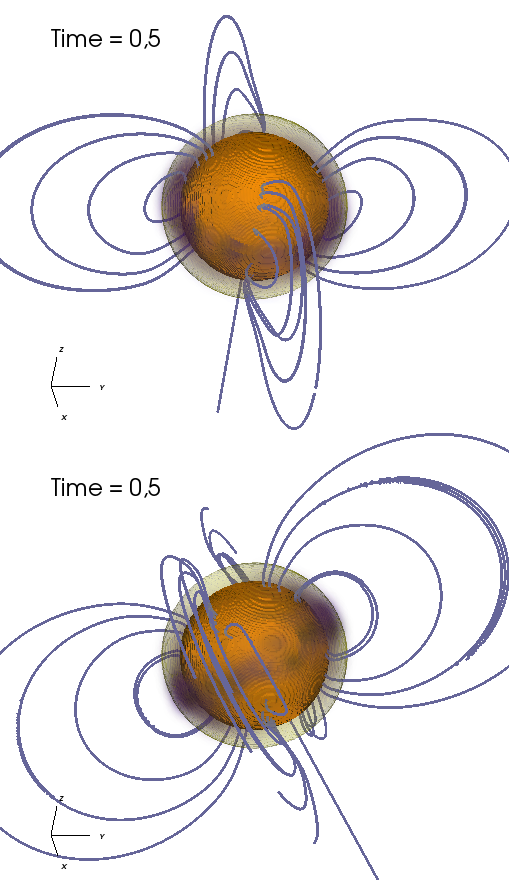} 
	\includegraphics[width=.32\linewidth]{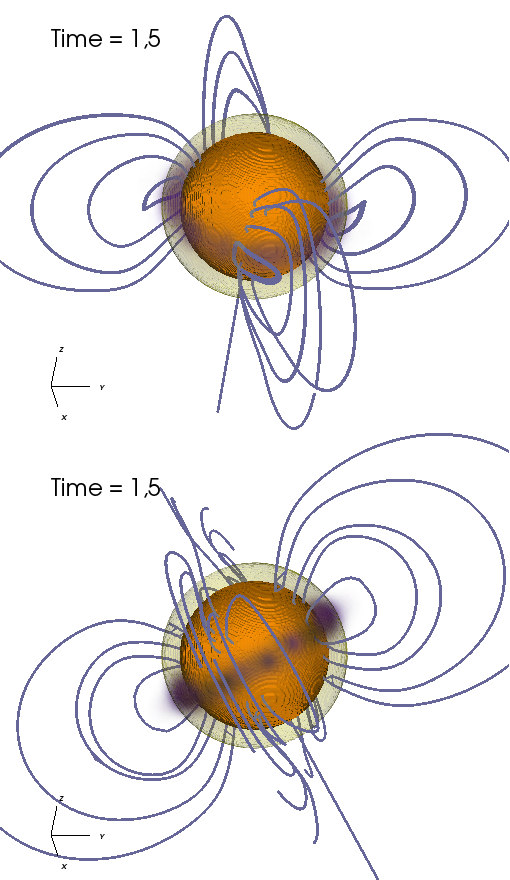} \\
	\caption{Evolution for the initially purely poloidal field case: 3d view of the magnetic field lines (blue). The orange and yellow surfaces indicate the interfaces $r=R_c$ and $r=R_\star$, respectively, while the purple color represents $j^2$. Top panels represent the case without tilt of the magnetic axis and bottom panels represent the case with a tilt given by the angles $\theta_{y}=\theta_{z}=\pi / 4$. Note that, due to technical reasons inherent to the visualization software {\em VisIt} used here, the poloidal field lines in the two cases are not necessarily the same.}
	\label{fig:poloidal_3d} 
\end{figure*}

\begin{figure}
    \centering
    \includegraphics[width=.8\linewidth]{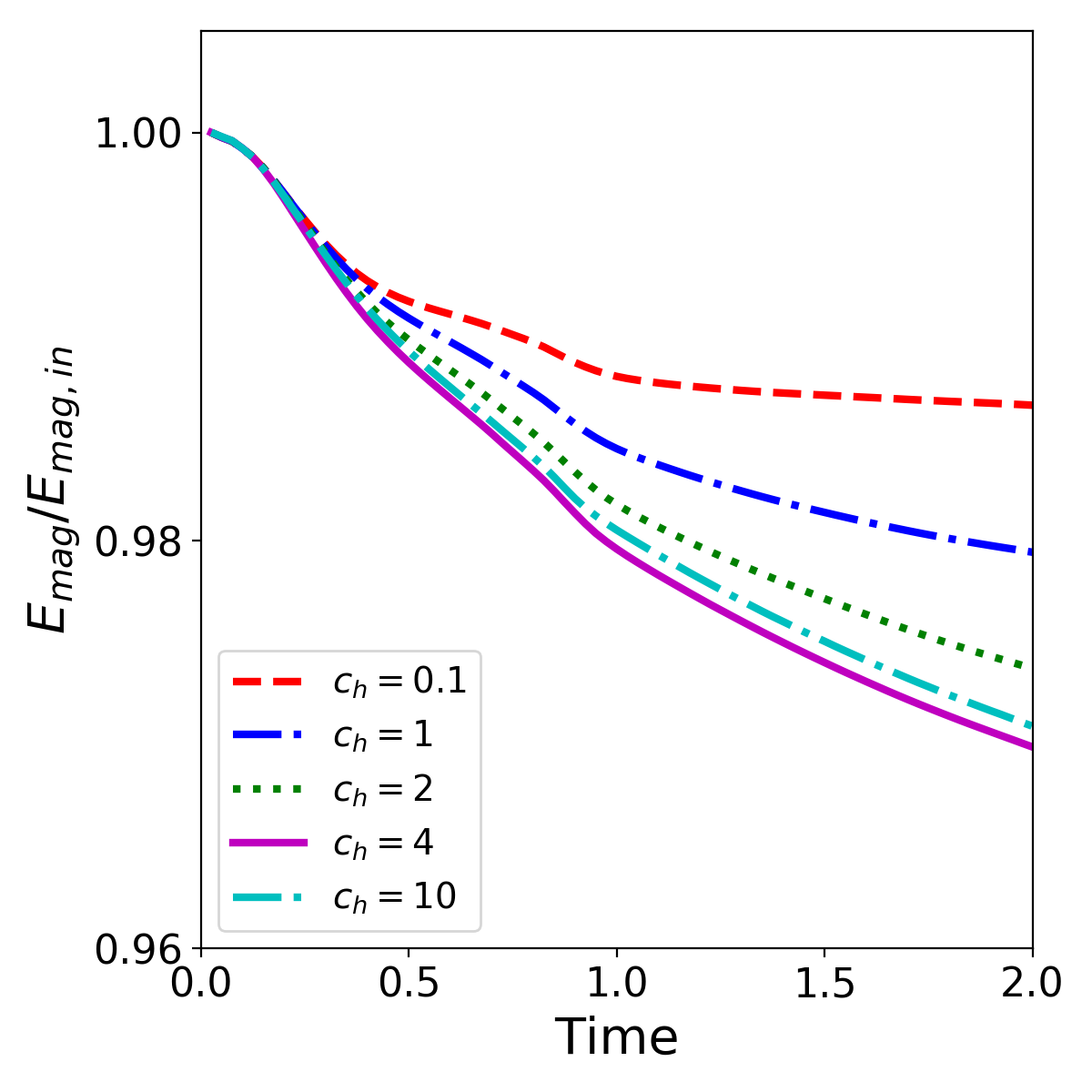} 
    \includegraphics[width=.8\linewidth]{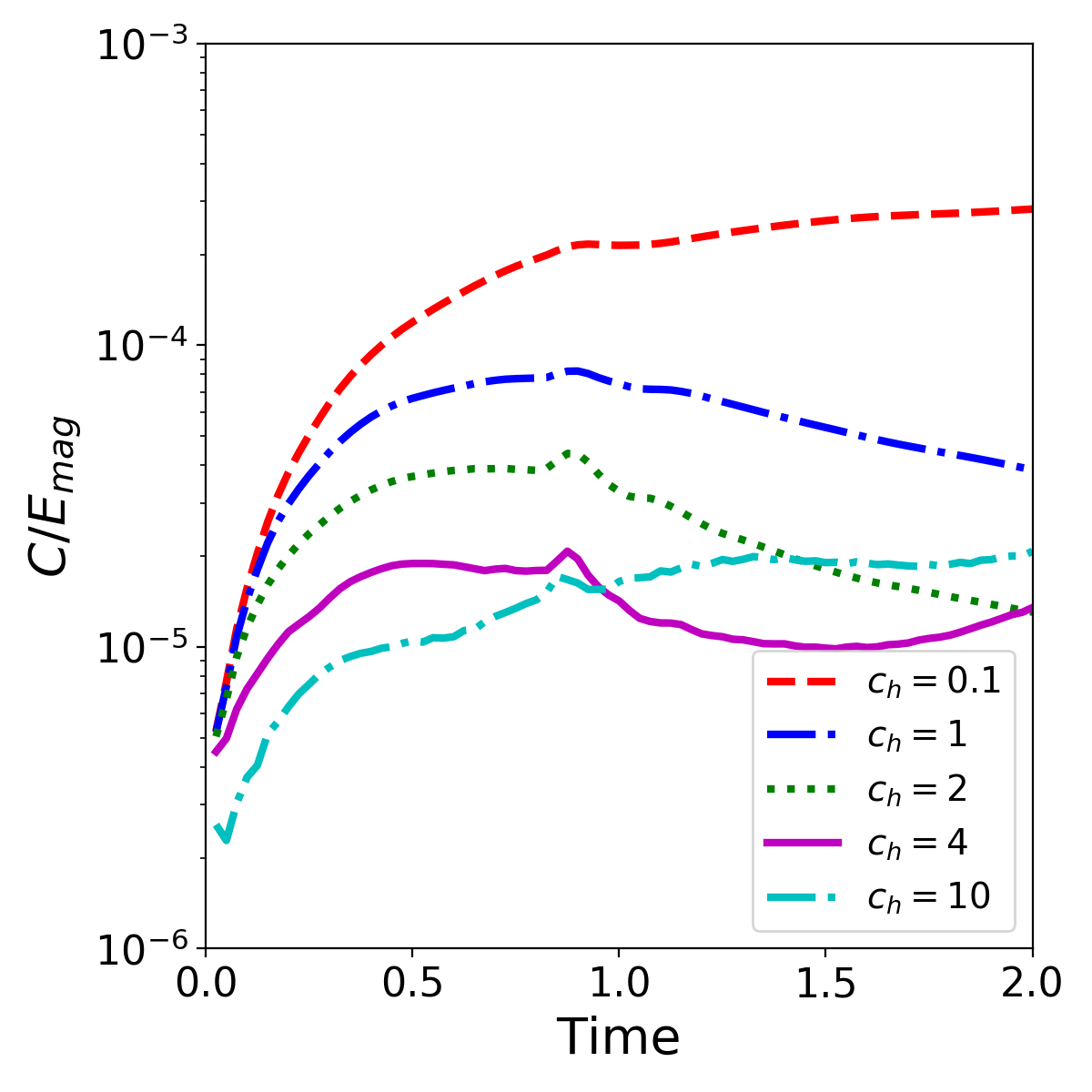} 
    \caption{Evolution of ${\cal E}_m$ (normalized by its initial value) and ${\cal C}/{\cal E}_m$ for different choices of $c_h$, for a non-dissipative case with only $f_h$ in the crust. Notice that, due to the non-linear character of the divergence cleaning equation, the constraint does not decay monotonically for $c_h \gtrsim 4$, affecting also the evolution of ${\cal E}_m$.} 
  \label{fig:poloidal_div} 
\end{figure}

We consider now a solution with an initially purely poloidal field, given by the following expressions for $r\leq R_\star$:

\begin{eqnarray}
  && B_r=B_0\pi^2\cos\theta \left(\frac{\sin \xi}{\xi^3} - \frac{\cos \xi}{\xi^2}\right)~,  \nonumber \\
  && B_\theta=B_0\pi^2\frac{\sin\theta}{2}\left(\frac{\sin \xi}{\xi^3}-\frac{\cos \xi}{\xi^2}- \frac{\sin \xi}{\xi}\right)~,\\
  && B_\varphi=0 ~.  \nonumber
\end{eqnarray}
where $\xi = \pi r/R_\star$. In the center of the star, $r\rightarrow 0$, the solution corresponding to a homogeneous field aligned with the magnetic axis $\vec{B} \rightarrow \frac{B_0\pi}{3}~\hat{z}$. Note that the poloidal components correspond to the solution in terms of spherical Bessel functions of test \S~\ref{sec:bessel}, which can be smoothly connected at the surface of the star $r=R_\star$ with a vacuum dipolar field, considered for $r > R_\star$:

\begin{eqnarray}
  && B_r=B_0\cos\theta\frac{R_\star^3}{r^3}~,  \nonumber\\
  && B_\theta=B_0\frac{\sin\theta}{2}\frac{R_\star^3}{r^3}~,\\
  && B_\varphi=0 ~.  \nonumber
\end{eqnarray}
As usual, we obtain the Cartesian components by using the spherical-to-Cartesian transformation shown in eqs.~(\ref{eq:sphtocart}).
Such initial configurations imply the presence of toroidal currents which circulate mainly in the core and support the dipolar magnetic field. Note that the currents are initially completely perpendicular to the magnetic field, i.e., far from a force-free configuration.

As before, we set $f_d=0$, keeping the same profile of $f_h$, and we set $f_a>0$ in the core. We have tried different values, and here we show $f_a=10$. The higher $f_a$, the faster the core dynamics, but the qualitative behavior is the same. In this case, the treatment of the region $r>R_\star$ is important, since our poloidal field extends outside the star. In order to compare the dynamics, we show two representative cases of the different treatments for the outer region: (i) {\em screened evolution}, with $f_a(r>R_\star)=0$, like in the toroidal test above, and (ii) {\em magneto-frictional coupling}, setting $f_a(r>R_\star)=100$ and a smooth transition between the interior and the exterior regions given by $f_a(R_m<r\leq R_\star) =  100(r-R_m)/(R_\star-R_m)$, with $R_m=9$.

In both cases, the boundary of our computational domain is far from the star's surface, at $L=30$, where open boundary conditions are applied just by extrapolating the interior solution. We have checked that, at this location, the boundary does not influence significantly the evolution within the domain (i.e., there are no spurious waves, line distortions or additional dissipation). In Fig.~\ref{fig:poloidal_initial} we show the initial configuration in the meridional plane, together with the base and the refined mesh (MR2) covering the shell. 

The evolution at different times (from left to right) is displayed in Fig.~\ref{fig:poloidal_evo} for the two cases (case (i) in the top row, case (ii) in the bottom row). As before, the simulation is 3D, but here we show the magnetic field lines and the toroidal field in a meridional cut, i.e. the $B_y$ component in the $x-z$ plane.

As expected, the initially poloidal dipolar field (supported by purely toroidal currents) develops a system of poloidal currents, which gives rise to a quadrupolar toroidal component in the crust, due to the Hall term. The corresponding toroidal field which rises has two opposite polarities in the core/inner crust and in the outer crust. The Hall drift in the crust drags the toroidal field close to the equator by means of the Burger-like dynamics analyzed above (\S~\ref{sec:hall_nl}).

In the core of both cases and in the magnetosphere of case (ii), the ambipolar term diffuses the crustal field out of the crust, and tries to bring the magnetic field into a force-free configuration, by removing the perpendicular currents $j_\perp = |\vec{j}-(\vec{j}\cdot\vec{B})\vec{B}/B^2|$. The evolution of the $j_\perp^2$ distribution in a meridional cut is displayed in color scales in Fig.~\ref{fig:poloidal_jperp_evo}. At the beginning, these currents are large in the core, since initially the magnetic field is purely poloidal and the current field is purely toroidal (initially, $j_\perp^2=j^2$). Therefore, the ambipolar term tends to straighten the poloidal magnetic field lines, thus reducing the underlying toroidal current. This process (i.e., transform both the initially purely poloidal magnetic field and purely toroidal current field into mixed poloidal-toroidal configuration) is the natural way to align the magnetic and the current fields, to gradually approach a force-free topology. Note that in the crust, where $f_a=0$, the perpendicular currents are not dissipated, consistently with what is expected.

The system of currents that is gradually developed is shown in Fig.~\ref{fig:poloidal_currents}, for the case (ii) at $t=0.5$. Currents in the poloidal projection are symmetric with respect to the equator, consistently with an anti-symmetric (mostly quadrupolar) toroidal magnetic field. In each quarter of the meridional plane, we can distinguish two vortices of currents (hereafter, loops for brevity) pointing to opposite directions, corresponding to toroidal fields of opposite polarities (blue/red in figure). Loops are separated by a layer in the inner crust with no toroidal field (white). All loops close at the equatorial plane and the magnetic axis. The current field for the case (i) is the same as case (ii), with the only difference that the external loops close with a very strong current sheet at the surface, instead of extending outside the star.

However, note that both cases shown here are very likely unphysical, since a smooth matching with a force-free solution outside the star is expected (i.e., currents and magnetic fields are parallel to each other everywhere to ensure $\vec{j}\times\vec{B}=0$, not like in our case (ii)), and no current sheet can be sustained at the star's surface (where actually the resistivity reaches its highest values, \cite{akgun18b}). Enforcing such matching is not trivial even in 2D polar coordinates (see the first simulations of this kind in \cite{akgun18b}), unless a potential (i.e., current-free) magnetospheric configuration is assumed, as usually done in 2D and 3D \cite{vigano12a,gourgouliatos18}. Therefore, in this test the system of currents and the magnetic dynamics are different from those previous works dealing with realistic cases \cite{vigano12a,gourgouliatos18}. In those cases, the boundary condition prevent currents from circulating in the outer layers of the star, giving rise to a different dynamics of the magnetic geometry. We defer for future works to explore further techniques and configurations to deal with a realistic star/magnetosphere matching.

Generally speaking, the interplay between the magnetic fields in the different regions depends on the difference in their timescales, which are set by the coefficients of the induction equation. If the timescales are very different at the two sides of a given interface (in contrast with what is shown in our toy model, where coefficients are of the same order across the interfaces), then discontinuities of the tangential magnetic field components can rise, producing a kink in the magnetic lines supported by current sheets flowing along the interface.

For the same reason, at the surface, in case (i), for which $f_a=0$ outside the star, kinks appear in the magnetic field lines (top panels of Fig.~\ref{fig:poloidal_evo}) and the currents close at the surface (see also the growing value of $j_\perp^2$ at the surface in the top panels of Fig.~\ref{fig:poloidal_jperp_evo}), while in case (ii) the magneto-frictional treatment employed avoids the appearance of the current sheet and allows the propagation of current and magnetic helicity outside the star (colors in the bottom panels of Fig.~\ref{fig:poloidal_evo}), accompanied by an inflation of the poloidal field lines (note that, in this particular toy model, the configuration outside is not force-free). We remark that we have explored the behavior for different profiles and magnitudes of $f_a$ and $f_h$, but here we only show the qualitative difference between the two mentioned cases. We refer to the next papers for a detailed discussion of the magneto-frictional coupling, the internal dynamics, and their physical meaning.

We also performed several numerical checks for this case. First, we run the same setups described above with a uniform grid with $N=400$, for which the evolved solutions are basically indistinguishable. This ensures that, for a given desired resolution of the shell, we can extend further away the domain by adding layers of refined meshes, decreasing the CPU time of the simulation by orders of magnitude compared to the uniform resolution case. The possibility of extending the domain very far away while keeping a high-resolution within the star will be particularly important in realistic cases, where a typical resolution of $O(10)$ m is required in the crust, followed by a smooth connection with a force-free magnetosphere that extends up to hundreds of stellar radii. The main differences obtained by changing the resolutions are negligible and result mainly from the possible discontinuities and surface geometry at the shell boundary.

Second, we check the conservation of energy and the constraint $\vec{\nabla}\cdot\vec{B}$ by evolving the same case studied above but only with the Hall term activated, which is the only one strictly not dissipative (i.e., we set $f_a=f_d=0$ everywhere, thus allowing the appearance of screening currents along both boundaries). The evolution of the integrated magnetic energy and the constraint deviation ${\cal C}$ are displayed as a function of time in Fig.~\ref{fig:poloidal_div}, where we have considered five different choices of the $c_h$ parameter (always maintaining $c_h=\kappa$) in the divergence cleaning equation. The magnetic energy is maintained at a level of $\lesssim 1-3\%$ at $t=2$, and the missing energy is numerically dissipated in the discontinuities at the equator and at the surfaces. On the other hand, the evolution of the solution, which is initially divergence-less, produces a rise of ${\cal C}$. After a short transient, it approaches a saturation level of ${\cal C} \ll 10^{-5}-10^{-3}~{\cal E}_m$, depending on the divergence damping timescales. As a matter of fact, the asymptotic value of ${\cal C}$ is inversely proportional to the value of $c_h$, if the latter has values O(1). As a compromise, we set $c_h=4$ in our simulations, which is a suitable value high enough to satisfy the constraints to a good accuracy but low enough not to restrain further the already small time-step. Note that, for this resolution and with this choice, ${\cal C} \ll 10^{-4}~ {\cal E}_m$ during all the simulations. The optimal values are problem-dependent, but we noted that with the chosen values of $c_h$ we obtain small deviations which have no impact on the evolution.

Last, we have also checked that the axial symmetry is preserved in our 3D Cartesian setup: no appreciable differences are seen among different meridional cuts. For the same reason, we also consider the same simulation with $N=200$, but tilting the magnetic axis of the initial condition by different angles around the $z$ and $y$-axis (by appropriately applying the transformation to coordinates and initial magnetic field components). Fig.~\ref{fig:poloidal_3d} shows the 3D view of magnetic field lines and current intensity squared, $j^2$, at different times, for two cases: with the magnetic axis along the $z$-axis (top panels, the standard initial data described above) and with tilts around $x$ and $y$ axes of $45^\circ$ and $45^\circ$, respectively (bottom panels). The axial symmetry is maintained in both cases. We found no significant differences between the two cases in the graphic visualizations and in the integrated quantities. The latter show negligible variations ($\lesssim 2\%$ for the chosen resolution) for different tilt angles at a given time. Note that such small differences can be also attributable to the numerical integration over the domain, which can give such differences due to the different orientation of the numerical cells.

This last set of tests shows that the Cartesian grid works in the same way, with or without symmetries around a given direction in the initial conditions.

\section{Conclusions}\label{sec:conclusions}

We have introduced a new finite-difference code to study a generalized 3D induction equation. Our code was automatically generated by using {\em Simflowny}, an open platform able to combine AMR optimal management, high-order accuracy both in space and time, modularity, scalability and user-friendliness.

We have explored three terms of a general induction equation: Ohmic, Hall and ambipolar, which are expected to play a role in the magnetic and thermal evolution of different regions and epochs of a neutron star. We considered HRSC methods, suitable for MHD, and explored several spatial reconstruction methods, combined with a fourth order RK temporal discretization scheme. These schemes are able to capture both smooth and non-smooth solutions, thus being really suitable for the Hall dynamics. 

We tested the code through a set of benchmark initial data, showing that different versions of well-known third-order and fifth-order schemes, belonging to the WENO and FDOC families, reproduce the Hall waves smooth analytical solutions with the expected accuracy. In other tests regarding the resistive and ambipolar terms, the convergence is limited to an order of about two, probably due to the numerical discretization of the curl operator used for the calculation of the current. Third-order schemes seem in general much more stable than fifth-order ones.

The novelty of our code lies in: (i) the use of a Cartesian 3D grid to solve at the same time all the regions of the star, including the magnetosphere; (ii) the efficient implementation, using high-order numerical schemes, on the infrastructure SAMRAI which allows for  high-scalability parallelization and  AMR~\cite{arbona18,palenzuela18}; (iii) the divergence cleaning method to ensure the constraint $\vec{\nabla}\cdot\vec{B}=0$; (iv) the generalization of the induction equation, including all possible terms that can act, either independently or combined.

We have checked that the 3D Cartesian grid works well in axisymmetric cases, and we performed an extended battery of checks to ensure that the well-known dynamics are correctly recovered. Also, although we only showed a simple illustrative case for simplicity, we underline that our code is able to manage an arbitrary number of dynamically refined layers, with an arbitrary refinement factor. These features might allow us to extend the external domain far away from the surface, and still resolve very accurately the most dynamical regions. This work paves the way for future 3D studies of the evolution of magnetic fields in neutron stars. We note that, in most tests presented here, we set the resistivity to zero, to challenge the numerical platform with an extreme case. Furthermore, we have not explored the parameter space of initial configurations of magnetic field, which could certainly be very different from the ones here assumed. 

Similarly, the crust-magnetosphere coupling has been briefly faced for one case, but it will be the main focus of a follow-up paper. Following the ideas proposed by \citep{akgun18}, we will explore the applications of the magneto-frictional method, in order to have a smooth connection with a magnetospheric configuration, and, notably, to study quantities like the rate of magnetic energy transferred to the magnetosphere, possibly related to the observed magnetar outburst activity.

In subsequent works, we aim at studying the long-term evolution for realistic profiles of the coefficients $f_d,f_h,f_a$. In particular, it will be interesting to explore the interior evolution under 3D ambipolar-dominated dynamics and the appearance of crustal small-scale magnetic structures, which could possibly be associated with the hotspots commonly inferred from the X-ray observations of isolated neutron stars. In order to do this, we will need to couple the temperature evolution, i.e., solving a parabolic PDE, taking into account the feedback between magnetic field and thermal evolution.

The platform {\em Simflowny} is easily adaptable to include the coupled fluid equations if needed, thus this code could actually be applied to other scenarios where Hall and ambipolar dynamics are relevant, including proto-planetary formation, molecular clouds, or the long-term magnetic evolution of other sources like exoplanets and white dwarfs.

\subsection*{Acknowledgments} 

We acknowledge support from the Spanish Ministry of Economy, Industry and Competitiveness grants AYA2016-80289-P and AYA2017-82089-ERC (AEI/FEDER, UE). CP also acknowledges support from the Spanish Ministry of Education and Science through a Ramon y Cajal grant. JAP acknowledges the Spanish MINECO/FEDER grant AYA2015-66899-C2-2-P, the grant of Generalitat Valenciana
PROMETEOII-2014-069. DM acknowledges support from Vicepresid\`encia i Conselleria d'Innovaci\'o, Recerca i Turisme del Govern de les Illes Balears. The work has been done within the PHAROS COST action CA16214. 

\bibliography{biblio}

\begin{thebibliography}{10}
\expandafter\ifx\csname url\endcsname\relax
  \def\url#1{\texttt{#1}}\fi
\expandafter\ifx\csname urlprefix\endcsname\relax\def\urlprefix{URL }\fi
\expandafter\ifx\csname href\endcsname\relax
  \def\href#1#2{#2} \def\path#1{#1}\fi

\bibitem{schunk77}
R.~W. {Schunk}, {Mathematical structure of transport equations for multispecies
  flows}, Reviews of Geophysics and Space Physics 15 (1977) 429--445.
\newblock \href {https://doi.org/10.1029/RG015i004p00429}
  {\path{doi:10.1029/RG015i004p00429}}.

\bibitem{witalis86}
E.~A. {Witalis}, {Hall magnetohydrodynamics and its applications to laboratory
  and cosmic plasma}, IEEE Transactions on Plasma Science 14 (1986) 842--848.
\newblock \href {https://doi.org/10.1109/TPS.1986.4316632}
  {\path{doi:10.1109/TPS.1986.4316632}}.

\bibitem{deng2001}
X.~H. {Deng}, H.~{Matsumoto}, {Rapid magnetic reconnection in the Earth's
  magnetosphere mediated by whistler waves}, Nature 410 (2001) 557--560.
\newblock \href {https://doi.org/10.1038/410557A0}
  {\path{doi:10.1038/410557A0}}.

\bibitem{mozer2002}
F.~S. {Mozer}, S.~D. {Bale}, T.~D. {Phan}, {Evidence of Diffusion Regions at a
  Subsolar Magnetopause Crossing}, Physical Review Letters 89~(1) (2002)
  015002.
\newblock \href {https://doi.org/10.1103/PhysRevLett.89.015002}
  {\path{doi:10.1103/PhysRevLett.89.015002}}.

\bibitem{bard18}
C.~{Bard}, J.~C. {Dorelli}, {On the role of system size in Hall MHD magnetic
  reconnection}, Physics of Plasmas 25~(2) (2018) 022103.
\newblock \href {http://arxiv.org/abs/1710.03612} {\path{arXiv:1710.03612}},
  \href {https://doi.org/10.1063/1.5010785} {\path{doi:10.1063/1.5010785}}.

\bibitem{kunz2004}
M.~W. {Kunz}, S.~A. {Balbus}, {Ambipolar diffusion in the magnetorotational
  instability}, \mnras 348 (2004) 355--360.
\newblock \href {http://arxiv.org/abs/astro-ph/0309707}
  {\path{arXiv:astro-ph/0309707}}, \href
  {https://doi.org/10.1111/j.1365-2966.2004.07383.x}
  {\path{doi:10.1111/j.1365-2966.2004.07383.x}}.

\bibitem{pandey2008}
B.~P. {Pandey}, M.~{Wardle}, {Hall magnetohydrodynamics of partially ionized
  plasmas}, \mnras 385 (2008) 2269--2278.
\newblock \href {http://arxiv.org/abs/0707.2688} {\path{arXiv:0707.2688}},
  \href {https://doi.org/10.1111/j.1365-2966.2008.12998.x}
  {\path{doi:10.1111/j.1365-2966.2008.12998.x}}.

\bibitem{bethune16}
W.~{B{\'e}thune}, G.~{Lesur}, J.~{Ferreira}, {Self-organisation in
  protoplanetary discs. Global, non-stratified Hall-MHD simulations}, \aap 589
  (2016) A87.
\newblock \href {http://arxiv.org/abs/1603.02475} {\path{arXiv:1603.02475}},
  \href {https://doi.org/10.1051/0004-6361/201527874}
  {\path{doi:10.1051/0004-6361/201527874}}.

\bibitem{huba91}
J.~D. {Huba}, {Theory and simulation of a high-frequency magnetic drift wave},
  Physics of Fluids B 3 (1991) 3217--3225.
\newblock \href {https://doi.org/10.1063/1.859752}
  {\path{doi:10.1063/1.859752}}.

\bibitem{huba03}
J.~D. {Huba}, {Hall Magnetohydrodynamics - A Tutorial}, in: {B{\"u}chner J.,
  Dum C. \& Scholer M.} (Ed.), Space Plasma Simulation, Vol. 615 of Lecture
  Notes in Physics, Berlin Springer Verlag, 2003, pp. 166--192.

\bibitem{khomenko2012}
E.~{Khomenko}, M.~{Collados}, {Heating of the Magnetized Solar Chromosphere by
  Partial Ionization Effects}, \apj 747 (2012) 87.
\newblock \href {http://arxiv.org/abs/1112.3374} {\path{arXiv:1112.3374}},
  \href {https://doi.org/10.1088/0004-637X/747/2/87}
  {\path{doi:10.1088/0004-637X/747/2/87}}.

\bibitem{soler15}
R.~{Soler}, M.~{Carbonell}, J.~L. {Ballester}, {On the Spatial Scales of Wave
  Heating in the Solar Chromosphere}, \apj 810 (2015) 146.
\newblock \href {http://arxiv.org/abs/1508.01497} {\path{arXiv:1508.01497}},
  \href {https://doi.org/10.1088/0004-637X/810/2/146}
  {\path{doi:10.1088/0004-637X/810/2/146}}.

\bibitem{fiedler92}
R.~A. {Fiedler}, T.~C. {Mouschovias}, {Ambipolar diffusion and star formation:
  Formation and contraction of axisymmetric cloud cores. I - Formulation of the
  problem and method of solution}, \apj 391 (1992) 199--219.
\newblock \href {https://doi.org/10.1086/171336} {\path{doi:10.1086/171336}}.

\bibitem{jones11}
A.~C. {Jones}, T.~P. {Downes}, {The Kelvin-Helmholtz instability in weakly
  ionized plasmas: ambipolar-dominated and Hall-dominated flows}, \mnras 418
  (2011) 390--400.
\newblock \href {http://arxiv.org/abs/1107.4241} {\path{arXiv:1107.4241}},
  \href {https://doi.org/10.1111/j.1365-2966.2011.19491.x}
  {\path{doi:10.1111/j.1365-2966.2011.19491.x}}.

\bibitem{gressel2015}
O.~{Gressel}, N.~J. {Turner}, R.~P. {Nelson}, C.~P. {McNally}, {Global
  Simulations of Protoplanetary Disks With Ohmic Resistivity and Ambipolar
  Diffusion}, \apj 801 (2015) 84.
\newblock \href {http://arxiv.org/abs/1501.05431} {\path{arXiv:1501.05431}},
  \href {https://doi.org/10.1088/0004-637X/801/2/84}
  {\path{doi:10.1088/0004-637X/801/2/84}}.

\bibitem{goldreich92}
P.~{Goldreich}, A.~{Reisenegger}, {Magnetic field decay in isolated neutron
  stars}, \apj 395 (1992) 250--258.
\newblock \href {https://doi.org/10.1086/171646} {\path{doi:10.1086/171646}}.

\bibitem{hollerbach02}
R.~{Hollerbach}, G.~{R{\"u}diger}, {The influence of Hall drift on the magnetic
  fields of neutron stars}, \mnras 337 (2002) 216--224.
\newblock \href {http://arxiv.org/abs/arXiv:astro-ph/0208312}
  {\path{arXiv:arXiv:astro-ph/0208312}}, \href
  {https://doi.org/10.1046/j.1365-8711.2002.05905.x}
  {\path{doi:10.1046/j.1365-8711.2002.05905.x}}.

\bibitem{hollerbach04}
R.~{Hollerbach}, G.~{R{\"u}diger}, {Hall drift in the stratified crusts of
  neutron stars}, \mnras 347 (2004) 1273--1278.
\newblock \href {https://doi.org/10.1111/j.1365-2966.2004.07307.x}
  {\path{doi:10.1111/j.1365-2966.2004.07307.x}}.

\bibitem{pons07b}
J.~A. {Pons}, U.~{Geppert}, {Magnetic field dissipation in neutron star crusts:
  from magnetars to isolated neutron stars}, \aap 470 (2007) 303--315.
\newblock \href {http://arxiv.org/abs/arXiv:astro-ph/0703267}
  {\path{arXiv:arXiv:astro-ph/0703267}}, \href
  {https://doi.org/10.1051/0004-6361:20077456}
  {\path{doi:10.1051/0004-6361:20077456}}.

\bibitem{pons09}
J.~A. {Pons}, J.~A. {Miralles}, U.~{Geppert}, {Magneto-thermal evolution of
  neutron stars}, \aap 496 (2009) 207--216.
\newblock \href {http://arxiv.org/abs/0812.3018} {\path{arXiv:0812.3018}},
  \href {https://doi.org/10.1051/0004-6361:200811229}
  {\path{doi:10.1051/0004-6361:200811229}}.

\bibitem{vigano12a}
D.~{Vigan{\`o}}, J.~A. {Pons}, J.~A. {Miralles}, {A new code for the
  Hall-driven magnetic evolution of neutron stars}, CoPhC 183 (2012)
  2042--2053.
\newblock \href {http://arxiv.org/abs/arXiv:astro-ph/1204.4707}
  {\path{arXiv:arXiv:astro-ph/1204.4707}}, \href
  {https://doi.org/10.1016/j.cpc.2012.04.029}
  {\path{doi:10.1016/j.cpc.2012.04.029}}.

\bibitem{vigano13}
D.~{Vigan{\`o}}, N.~{Rea}, J.~A. {Pons}, R.~{Perna}, D.~N. {Aguilera}, J.~A.
  {Miralles}, {Unifying the observational diversity of isolated neutron stars
  via magneto-thermal evolution models}, \mnras 434 (2013) 123--141.
\newblock \href {http://arxiv.org/abs/1306.2156} {\path{arXiv:1306.2156}},
  \href {https://doi.org/10.1093/mnras/stt1008}
  {\path{doi:10.1093/mnras/stt1008}}.

\bibitem{geppert14}
U.~{Geppert}, D.~{Vigan{\`o}}, {Creation of magnetic spots at the neutron star
  surface}, \mnras 444 (2014) 3198--3208.
\newblock \href {http://arxiv.org/abs/1408.3833} {\path{arXiv:1408.3833}},
  \href {https://doi.org/10.1093/mnras/stu1675}
  {\path{doi:10.1093/mnras/stu1675}}.

\bibitem{gourgouliatos13}
K.~N. {Gourgouliatos}, A.~{Cumming}, A.~{Reisenegger}, C.~{Armaza},
  M.~{Lyutikov}, J.~A. {Valdivia}, {Hall equilibria with toroidal and poloidal
  fields: application to neutron stars}, \mnras 434 (2013) 2480--2490.
\newblock \href {http://arxiv.org/abs/1305.6269} {\path{arXiv:1305.6269}},
  \href {https://doi.org/10.1093/mnras/stt1195}
  {\path{doi:10.1093/mnras/stt1195}}.

\bibitem{gourgouliatos14a}
K.~N. {Gourgouliatos}, A.~{Cumming}, {Hall effect in neutron star crusts:
  evolution, endpoint and dependence on initial conditions}, \mnras 438 (2014)
  1618--1629.
\newblock \href {http://arxiv.org/abs/1311.7004} {\path{arXiv:1311.7004}},
  \href {https://doi.org/10.1093/mnras/stt2300}
  {\path{doi:10.1093/mnras/stt2300}}.

\bibitem{gourgouliatos14b}
K.~N. {Gourgouliatos}, A.~{Cumming}, {Hall Attractor in Axially Symmetric
  Magnetic Fields in Neutron Star Crusts}, Physical Review Letters 112~(17)
  (2014) 171101.
\newblock \href {http://arxiv.org/abs/1311.7345} {\path{arXiv:1311.7345}},
  \href {https://doi.org/10.1103/PhysRevLett.112.171101}
  {\path{doi:10.1103/PhysRevLett.112.171101}}.

\bibitem{wood2015}
T.~S. {Wood}, R.~{Hollerbach}, {Three Dimensional Simulation of the Magnetic
  Stress in a Neutron Star Crust}, Physical Review Letters 114~(19) (2015)
  191101.
\newblock \href {http://arxiv.org/abs/1501.05149} {\path{arXiv:1501.05149}},
  \href {https://doi.org/10.1103/PhysRevLett.114.191101}
  {\path{doi:10.1103/PhysRevLett.114.191101}}.

\bibitem{gourgouliatos18}
K.~N. {Gourgouliatos}, R.~{Hollerbach}, {Magnetic Axis Drift and Magnetic Spot
  Formation in Neutron Stars with Toroidal Fields}, \apj 852 (2018) 21.
\newblock \href {http://arxiv.org/abs/1710.01338} {\path{arXiv:1710.01338}},
  \href {https://doi.org/10.3847/1538-4357/aa9d93}
  {\path{doi:10.3847/1538-4357/aa9d93}}.

\bibitem{elfritz16}
J.~G. {Elfritz}, J.~A. {Pons}, N.~{Rea}, K.~{Glampedakis}, D.~{Vigan{\`o}},
  {Simulated magnetic field expulsion in neutron star cores}, \mnras 456 (2016)
  4461--4474.
\newblock \href {http://arxiv.org/abs/1512.07151} {\path{arXiv:1512.07151}},
  \href {https://doi.org/10.1093/mnras/stv2963}
  {\path{doi:10.1093/mnras/stv2963}}.

\bibitem{castillo17}
F.~{Castillo}, A.~{Reisenegger}, J.~A. {Valdivia}, {Magnetic field evolution
  and equilibrium configurations in neutron star cores: the effect of ambipolar
  diffusion}, \mnras 471 (2017) 507--522.
\newblock \href {http://arxiv.org/abs/1705.10020} {\path{arXiv:1705.10020}},
  \href {https://doi.org/10.1093/mnras/stx1604}
  {\path{doi:10.1093/mnras/stx1604}}.

\bibitem{gusakov17}
M.~E. {Gusakov}, E.~M. {Kantor}, D.~D. {Ofengeim}, {Evolution of the magnetic
  field in neutron stars}, \prd 96~(10) (2017) 103012.
\newblock \href {http://arxiv.org/abs/1705.00508} {\path{arXiv:1705.00508}},
  \href {https://doi.org/10.1103/PhysRevD.96.103012}
  {\path{doi:10.1103/PhysRevD.96.103012}}.

\bibitem{passamonti17a}
A.~{Passamonti}, T.~{Akg{\"u}n}, J.~A. {Pons}, J.~A. {Miralles}, {On the
  magnetic field evolution time-scale in superconducting neutron star cores},
  \mnras 469 (2017) 4979--4984.
\newblock \href {http://arxiv.org/abs/1704.02016} {\path{arXiv:1704.02016}},
  \href {https://doi.org/10.1093/mnras/stx1192}
  {\path{doi:10.1093/mnras/stx1192}}.

\bibitem{passamonti17b}
A.~{Passamonti}, T.~{Akg{\"u}n}, J.~A. {Pons}, J.~A. {Miralles}, {The relevance
  of ambipolar diffusion for neutron star evolution}, \mnras 465 (2017)
  3416--3428.
\newblock \href {http://arxiv.org/abs/1608.00001} {\path{arXiv:1608.00001}},
  \href {https://doi.org/10.1093/mnras/stw2936}
  {\path{doi:10.1093/mnras/stw2936}}.

\bibitem{kantor18}
E.~M. {Kantor}, M.~E. {Gusakov}, {A note on the ambipolar diffusion in
  superfluid neutron stars}, \mnras 473 (2018) 4272--4277.
\newblock \href {http://arxiv.org/abs/1703.09216} {\path{arXiv:1703.09216}},
  \href {https://doi.org/10.1093/mnras/stx2682}
  {\path{doi:10.1093/mnras/stx2682}}.

\bibitem{bransgrove18}
A.~{Bransgrove}, Y.~{Levin}, A.~{Beloborodov}, {Magnetic field evolution of
  neutron stars - I. Basic formalism, numerical techniques and first results},
  \mnras 473 (2018) 2771--2790.
\newblock \href {http://arxiv.org/abs/1709.09167} {\path{arXiv:1709.09167}},
  \href {https://doi.org/10.1093/mnras/stx2508}
  {\path{doi:10.1093/mnras/stx2508}}.

\bibitem{ofengeim18}
D.~D. {Ofengeim}, M.~E. {Gusakov}, {Fast magnetic field evolution in neutron
  stars: The key role of magnetically induced fluid motions in the core}, \prd
  98~(4) (2018) 043007.
\newblock \href {http://arxiv.org/abs/1805.03956} {\path{arXiv:1805.03956}},
  \href {https://doi.org/10.1103/PhysRevD.98.043007}
  {\path{doi:10.1103/PhysRevD.98.043007}}.

\bibitem{arbona13}
A.~{Arbona}, A.~{Artigues}, C.~{Bona-Casas}, J.~{Mass{\'o}}, B.~{Mi{\~n}ano},
  A.~{Rigo}, M.~{Trias}, C.~{Bona}, {Simflowny: A general-purpose platform for
  the management of physical models and simulation problems}, Computer Physics
  Communications 184 (2013) 2321--2331.
\newblock \href {https://doi.org/10.1016/j.cpc.2013.04.012}
  {\path{doi:10.1016/j.cpc.2013.04.012}}.

\bibitem{arbona18}
A.~{Arbona}, B.~{Minano}, A.~{Rigo}, B.~C., C.~{Palenzuela}, A.~{Artigues},
  C.~{Bona-Casas}, J.~{Massó}, {Simflowny 2: An upgraded platform for
  scientific modelling and simulation}, CoPhC 231.
\newblock \href {https://doi.org/10.1016/j.cpc.2018.03.015}
  {\path{doi:10.1016/j.cpc.2018.03.015}}.

\bibitem{hornung02}
R.~D. Hornung, S.~R. Kohn, \href{http://dx.doi.org/10.1002/cpe.652}{Managing
  application complexity in the samrai object-oriented framework}, Concurrency
  and Computation: Practice and Experience 14~(5) (2002) 347--368.
\newblock \href {https://doi.org/10.1002/cpe.652} {\path{doi:10.1002/cpe.652}}.
\newline\urlprefix\url{http://dx.doi.org/10.1002/cpe.652}

\bibitem{gunney16}
B.~T. Gunney, R.~W. Anderson,
  \href{http://www.sciencedirect.com/science/article/pii/S0743731515002129}{Advances
  in patch-based adaptive mesh refinement scalability}, Journal of Parallel and
  Distributed Computing 89 (2016) 65 -- 84.
\newblock \href {https://doi.org/https://doi.org/10.1016/j.jpdc.2015.11.005}
  {\path{doi:https://doi.org/10.1016/j.jpdc.2015.11.005}}.
\newline\urlprefix\url{http://www.sciencedirect.com/science/article/pii/S0743731515002129}

\bibitem{geppert91}
U.~{Geppert}, H.-J. {Wiebicke}, {Amplification of neutron star magnetic fields
  by thermoelectric effects. I - General formalism}, AAPS 87 (1991) 217--228.

\bibitem{vainshtein00}
S.~I. {Vainshtein}, S.~M. {Chitre}, A.~V. {Olinto}, {Rapid dissipation of
  magnetic fields due to the Hall current}, \pre 61 (2000) 4422--4430.
\newblock \href {http://arxiv.org/abs/arXiv:astro-ph/9911386}
  {\path{arXiv:arXiv:astro-ph/9911386}}, \href
  {https://doi.org/10.1103/PhysRevE.61.4422}
  {\path{doi:10.1103/PhysRevE.61.4422}}.

\bibitem{roumeliotis94}
G.~{Roumeliotis}, P.~A. {Sturrock}, S.~K. {Antiochos}, {A Numerical Study of
  the Sudden Eruption of Sheared Magnetic Fields}, \apj 423 (1994) 847--+.
\newblock \href {https://doi.org/10.1086/173862} {\path{doi:10.1086/173862}}.

\bibitem{vigano11}
D.~{Vigan{\`o}}, J.~A. {Pons}, J.~A. {Miralles}, {Force-free twisted
  magnetospheres of neutron stars}, \aap 533 (2011) A125.
\newblock \href {http://arxiv.org/abs/1106.5934} {\path{arXiv:1106.5934}},
  \href {https://doi.org/10.1051/0004-6361/201117105}
  {\path{doi:10.1051/0004-6361/201117105}}.

\bibitem{helliwell65}
R.~A. {Helliwell}, {Whistlers and Related Ionospheric Phenomena}, 1965.

\bibitem{nunn74}
D.~{Nunn}, {A self-consistent theory of triggered VLF emissions}, Planss 22
  (1974) 349--378.
\newblock \href {https://doi.org/10.1016/0032-0633(74)90070-1}
  {\path{doi:10.1016/0032-0633(74)90070-1}}.

\bibitem{toth00}
G.~{T{\'o}th}, Journal of Computational Physics 161 (2000) 605--652.
\newblock \href {https://doi.org/10.1006/jcph.2000.6519}
  {\path{doi:10.1006/jcph.2000.6519}}.

\bibitem{dedner02}
A.~{Dedner}, F.~{Kemm}, D.~{Kr{\"o}ner}, C.-D. {Munz}, T.~{Schnitzer},
  M.~{Wesenberg}, {Hyperbolic Divergence Cleaning for the MHD Equations},
  Journal of Computational Physics 175 (2002) 645--673.
\newblock \href {https://doi.org/10.1006/jcph.2001.6961}
  {\path{doi:10.1006/jcph.2001.6961}}.

\bibitem{toro97}
E.~Toro, \href{https://books.google.es/books?id=6QFAAQAAIAAJ}{Riemann Solvers
  and Numerical Methods for Fluid Dynamics: A Practical Introduction},
  Springer, 1997.
\newline\urlprefix\url{https://books.google.es/books?id=6QFAAQAAIAAJ}

\bibitem{colella84}
P.~{Colella}, P.~R. {Woodward}, {The Piecewise Parabolic Method (PPM) for
  Gas-Dynamical Simulations}, Journal of Computational Physics 54 (1984)
  174--201.
\newblock \href {https://doi.org/10.1016/0021-9991(84)90143-8}
  {\path{doi:10.1016/0021-9991(84)90143-8}}.

\bibitem{suresh97}
A.~Suresh, H.~Huynh,
  \href{http://www.sciencedirect.com/science/article/pii/S0021999197957454}{Accurate
  monotonicity-preserving schemes with runge–kutta time stepping}, Journal of
  Computational Physics 136~(1) (1997) 83 -- 99.
\newblock \href {https://doi.org/https://doi.org/10.1006/jcph.1997.5745}
  {\path{doi:https://doi.org/10.1006/jcph.1997.5745}}.
\newline\urlprefix\url{http://www.sciencedirect.com/science/article/pii/S0021999197957454}

\bibitem{bona09}
C.~{Bona}, C.~{Bona-Casas}, J.~{Terradas}, {Linear high-resolution schemes for
  hyperbolic conservation laws: TVB numerical evidence}, Journal of
  Computational Physics 228 (2009) 2266--2281.
\newblock \href {http://arxiv.org/abs/0810.2185} {\path{arXiv:0810.2185}},
  \href {https://doi.org/10.1016/j.jcp.2008.12.010}
  {\path{doi:10.1016/j.jcp.2008.12.010}}.

\bibitem{jiang96}
G.-S. Jiang, C.-W. Shu,
  \href{http://www.sciencedirect.com/science/article/pii/S0021999196901308}{Efficient
  implementation of weighted eno schemes}, Journal of Computational Physics
  126~(1) (1996) 202 -- 228.
\newblock \href {https://doi.org/https://doi.org/10.1006/jcph.1996.0130}
  {\path{doi:https://doi.org/10.1006/jcph.1996.0130}}.
\newline\urlprefix\url{http://www.sciencedirect.com/science/article/pii/S0021999196901308}

\bibitem{shu98}
C.-W. Shu, \href{https://doi.org/10.1007/BFb0096355}{Essentially
  non-oscillatory and weighted essentially non-oscillatory schemes for
  hyperbolic conservation laws}, Springer Berlin Heidelberg, Berlin,
  Heidelberg, 1998, pp. 325--432.
\newblock \href {https://doi.org/10.1007/BFb0096355}
  {\path{doi:10.1007/BFb0096355}}.
\newline\urlprefix\url{https://doi.org/10.1007/BFb0096355}

\bibitem{palenzuela18}
C.~{Palenzuela}, B.~{Mi{\~n}ano}, D.~{Vigan{\`o}}, A.~{Arbona},
  C.~{Bona-Casas}, A.~{Rigo}, M.~{Bezares}, C.~{Bona}, J.~{Mass{\'o}}, {A
  Simflowny-based finite-difference code for high-performance computing in
  numerical relativity}, Classical and Quantum Gravity 35~(18) (2018) 185007.
\newblock \href {http://arxiv.org/abs/1806.04182} {\path{arXiv:1806.04182}},
  \href {https://doi.org/10.1088/1361-6382/aad7f6}
  {\path{doi:10.1088/1361-6382/aad7f6}}.

\bibitem{gonzalezmorales2018}
P.~A. {Gonz{\'a}lez-Morales}, E.~{Khomenko}, T.~P. {Downes}, A.~{de Vicente},
  {MHDSTS: a new explicit numerical scheme for simulations of partially ionised
  solar plasma}, \aap 615 (2018) A67.
\newblock \href {http://arxiv.org/abs/1803.04891} {\path{arXiv:1803.04891}},
  \href {https://doi.org/10.1051/0004-6361/201731916}
  {\path{doi:10.1051/0004-6361/201731916}}.

\bibitem{osullivan06}
S.~{O'Sullivan}, T.~P. {Downes}, {An explicit scheme for multifluid
  magnetohydrodynamics}, \mnras 366 (2006) 1329--1336.
\newblock \href {http://arxiv.org/abs/astro-ph/0511478}
  {\path{arXiv:astro-ph/0511478}}, \href
  {https://doi.org/10.1111/j.1365-2966.2005.09898.x}
  {\path{doi:10.1111/j.1365-2966.2005.09898.x}}.

\bibitem{Calabrese:2004}
G.~{Calabrese}, L.~{Lehner}, O.~{Reula}, O.~{Sarbach}, M.~{Tiglio}, {Summation
  by parts and dissipation for domains with excised regions}, Classical and
  Quantum Gravity 21 (2004) 5735--5757.
\newblock \href {http://arxiv.org/abs/gr-qc/0308007}
  {\path{arXiv:gr-qc/0308007}}, \href
  {https://doi.org/10.1088/0264-9381/21/24/004}
  {\path{doi:10.1088/0264-9381/21/24/004}}.

\bibitem{yamaleev09}
N.~K. Yamaleev, M.~H. Carpenter,
  \href{http://www.sciencedirect.com/science/article/pii/S002199910900014X}{Third-order
  energy stable weno scheme}, Journal of Computational Physics 228~(8) (2009)
  3025 -- 3047.
\newblock \href {https://doi.org/https://doi.org/10.1016/j.jcp.2009.01.011}
  {\path{doi:https://doi.org/10.1016/j.jcp.2009.01.011}}.
\newline\urlprefix\url{http://www.sciencedirect.com/science/article/pii/S002199910900014X}

\bibitem{barenblatt52}
G.~I. {Barenblatt}, {On some unsteady fluid and gas motions in a porous
  medium.}, Prikladnaya Matematika i Mekhanika 16~(1).

\bibitem{pattle59}
R.~E. {Pattle}, {DIFFUSION FROM AN INSTANTANEOUS POINT SOURCE WITH A
  CONCENTRATION-DEPENDENT COEFFICIENT}, The Quarterly Journal of Mechanics and
  Applied Mathematics 12~(4).

\bibitem{vigano12b}
D.~{Vigan{\`o}}, J.~A. {Pons}, {Central compact objects and the hidden magnetic
  field scenario}, \mnras 425 (2012) 2487--2492.
\newblock \href {http://arxiv.org/abs/1206.2014} {\path{arXiv:1206.2014}},
  \href {https://doi.org/10.1111/j.1365-2966.2012.21679.x}
  {\path{doi:10.1111/j.1365-2966.2012.21679.x}}.

\bibitem{akgun18b}
T.~{Akg{\"u}n}, P.~{Cerd{\'a}-Dur{\'a}n}, J.~A. {Miralles}, J.~A. {Pons},
  {Crust-magnetosphere coupling during magnetar evolution and implications for
  the surface temperature}, \mnras 481 (2018) 5331--5338.
\newblock \href {http://arxiv.org/abs/1807.09021} {\path{arXiv:1807.09021}},
  \href {https://doi.org/10.1093/mnras/sty2669}
  {\path{doi:10.1093/mnras/sty2669}}.

\bibitem{akgun18}
T.~{Akg{\"u}n}, P.~{Cerd{\'a}-Dur{\'a}n}, J.~A. {Miralles}, J.~A. {Pons}, {The
  force-free twisted magnetosphere of a neutron star - II. Degeneracies of the
  Grad-Shafranov equation}, \mnras 474 (2018) 625--635.
\newblock \href {http://arxiv.org/abs/1709.10044} {\path{arXiv:1709.10044}},
  \href {https://doi.org/10.1093/mnras/stx2814}
  {\path{doi:10.1093/mnras/stx2814}}.

\end{thebibliography}
\bibliographystyle{elsarticle-num}

\appendix

\section{Discretized current}\label{app:j}

We tried with different prescription of the current. Note that its definition, eq.~(\ref{eq:current_mhd}), can be discretized in two ways: either by a direct finite difference in the two direction, or applying the Stokes theorem and giving some prescription for the circulation $C_{kl}$ of the magnetic field (where $(k,l)$ is the plane perpendicular to the current component $J_m$ that we are calculating, and $i,j$ identify their correspondent discretized position along the two directions). The circulation can be chosen by considering the line integral along the square centered in the point $i,j$, with sides given by $2\Delta x_i$, $2\Delta x_j$, i.e., the square delimited by the 8 closest points of the grid in the plane: the values at the center and corners of the four sides. In general, the line integral can be evaluated as:

\begin{eqnarray}
 && J_m = \epsilon^{klm} C_{kl} \\
 && C_{kl} = \{ w_s [B_l(i+1,j-1) - B_l(i-1,j-1)] \nonumber \\
 && + w_c [B_l(i+1,j) - B_l(i-1,j)] \nonumber \\
 && + w_s [B_l(i+1,j+1) - B_l(i+1,j-1)]\} \nonumber \\
 && - \{w_s [B_k(i-1,j+1) - B_k(i-1,j-1)] \nonumber \\
 && + w_c [B_k(i,j+1) - B_k(i,j-1)] \nonumber \\
 && + w_s [B_k(i+1,j+1) - B_k(i+1,j-1)]\} \nonumber 
\end{eqnarray}
$w_s$ and $w_c$ the non-negative weights to the side and central neighbors, respectively, so that $2w_s + w_c=1$. We tested different combination of weights within the range $w_c \in [0,1]$. When comparing the errors in the following tests, we noted that the convergence order does not depend on the choice of such parameters, and the relative errors are minimized for the choice $w_c=1, w_s=0$, which is actually the purely finite difference scheme with no correction coming from the corner neighbors. This can be explained by the fact that the inclusion of the values of the corner means that we are including a correction in one direction (let us say, $k$), when evaluating the finite difference along the perpendicular direction ($l$). In this sense, the form of the curl operator is intrinsically different from a $n^{\rm th}$-order derivative operator acting in one direction only: for the latter, extending the stencil in the direction in a proper way improves the accuracy. We also note that enlarging the number of stencil points for the finite difference version ($w_c=1$) makes the code very unstable.

A potentially interesting alternative is to evolve the current density $\vec{J}$ independently, instead of defining it by eq.~(\ref{eq:current_mhd}). By taking the curl to eq.~\ref{eq:ind_general}, we obtain the following equation in the non-conservative form

\begin{equation}
   \frac{\partial \vec{J}}{\partial t} - \vec{\nabla}(\vec{\nabla}\cdot\vec{E}) = - \nabla^2 \vec{E}
\end{equation}
where $\vec{E}$ is given by eq.~\ref{eq:efield_general}. The flux term contains first-order derivatives, and the source term has second-order derivatives of the variables. This makes the resulting code very unstable with our HRSC methods, thus making this formulation not apt for the numerical techniques here employed.

\section{Some extensions to the characteristic analysis of the system}\label{app:characteristic}

\subsection{Particular solutions in Cartesian coordinates}\label{app:whistler}

If we particularize the general characteristic solutions \eqref{eigenvalues}-\eqref{eigenvectors} to a Cartesian coordinate system, by assuming $B_{ok}\neq0$, we may chose:
 \be
 \vec{k}= \frac{k}{\sqrt{2}} (\hat{x}+\hat{y}) \quad \text{,} \quad \vec{B}_o = B_o \hat{x} 
 \ee
(i.e. an angle $45^\circ$ between $\vec{k}$ and $\vec{B}_o$)\footnote{Notice there is no loss of generality by adopting this particular choice and it could be generalized easily, as long as $\vec{k}\cdot\vec{B}_o \neq 0$.}. 
A straightforward calculation then yields,
 \begin{eqnarray*}
 \vec{B}_{1}^{\pm} &=&  2f_h (\hat{x} - \hat{y}) + \left[ f_a B_o \mp \sqrt{f_{a}^2 B_{o}^2 - 8f_{h}^2} \right]  \hat{z} \\ 
 i \omega^{\pm} &=& k^2 \left\lbrace f_d + \frac{3}{4} f_a B_{o}^2 \pm \frac{B_o}{4} \sqrt{f_{a}^2 B_{o}^2 - 8f_{h}^2} \right\rbrace 
 \end{eqnarray*}
which in the ideal case, $f_d = f_a = 0$ (selecting the ``+'' wave-mode), we can re-write as:
 \be
  \vec{B} = B_o \, \hat{x} + B_1 \left[ \hat{x} -\hat{y} + i \sqrt{2}\, \hat{z}\right] e^{i\left( \frac{k}{\sqrt{2}}(x+y) - \omega t \right) } \nonumber
 \ee
where $\omega = f_h k^2 B_o /\sqrt{2}$. This is equivalent to the initial data of the whistler test eq.~(\ref{whistler_data}), by setting $t=0$, $k_x \equiv k/\sqrt{2}$; $v_{\omega} \equiv \omega/k_x$, and taking just the real part. The solution consists of a circularly polarized wave in the plane perpendicular to $\hat{x}+\hat{y}$.

Another interesting solution can be found by looking at the case $\vec{k} \parallel \vec{B}_o$ (i.e. $\vec{B}_{op} = 0$). Thus,
\be
 \vec{k}= k \hat{x} \quad \text{,} \quad \vec{B}_o = B_o \hat{x} 
\ee
leading to the damped whistler wave described by,
\be
  \vec{B} = B_o \, \hat{x} + B_1 \left[ \hat{y} + i \hat{z}\right] e^{i\left( k x -  \omega_h t \right) } e^{-t/\tau_d } \nonumber
 \ee
with $\omega_h = f_h k^2 B_{ok}$ being the whistler frequency of the propagation and  $1/\tau_d = k^2 \left( f_d + f_a B_{o}^2 \right)$ controlling the parabolic decay.

\subsection{Non-homogeneous coefficients}\label{app:nonhomogeneous}
If we relax the assumption of high-frequency limit, we can consider more general solutions to the linearized problem by studying equations with non-negligible gradients of the coefficients:
$f_d = f_d (\vec{x}) $, $f_a = f_a (\vec{x}) $ or $f_h = f_h (\vec{x}) $.

Let us define the gradient as, $\vec{H}:=\vec{\nabla} f_h$. Then, the general equation we need to solve is:
 \begin{eqnarray}
i \omega \vec{B}_1 &=& \omega_d \vec{B}_{1p} - \omega_h \vec{B}_{1q} + f_a k^2 (\vec{B}_o \cdot \vec{B}_{1q} ) \vec{B}_{oq} \nonumber\\
&& + ik \left[ (\vec{H} \cdot \vec{B}_{o} ) \vec{B}_{1q} - (\vec{H} \cdot \vec{B}_{1q} ) \vec{B}_{o}\right] 
\end{eqnarray}
 where we have used again that $\omega_h = f_h k^2 B_{ok}$ and $\omega_d = k^2 \left( f_d + f_a B_{o}^2 \right)$.
 This leads to a complicated dispersion relation for the physical modes of the system, namely:
 \be
\left( X + i k H_{oq} \right) \left( X + f_a k^2 B_{op}^2 \right) +  Y \left(Y - ik H_{op} \right) = 0 \nonumber
 \ee
 where $X:=i\omega -\omega_d$, $Y:= \omega_h - i k H_k B_{ok}$ and $H_{op/q}\equiv \vec{H}\cdot \vec{B}_{op/q}$. Note it is a quadratic system for X, but with complex coefficients. To our knowledge, there is no trivial generic solution.
 
 We shall then particularize the above system to some interesting cases within the Hall MHD approximation ($f_d = f_a = 0$), setting the propagation direction orthogonal to both $\vec{B}_o$ and $\vec{H}$. In this case the dispersion relation reduces to $\omega (\omega +  k H_{oq})= 0$, thus recovering the so-called Hall drift mode (see \cite{huba91}), which can be written,
 \be
 \vec{B}_1 = \vec{H}_q \quad \text{,} \quad \omega = -k H_{oq}
 \ee

 It might be relevant to note that within Hall MHD there are no allowed solutions that propagate along the $f_h$ gradient, i.e. $\vec{k} \parallel \vec{H}$.

\end{document}